\documentclass[fleqn,usenatbib]{mnras}

\usepackage{newtxtext,newtxmath}
\usepackage{verbatim}
\usepackage[T1]{fontenc}

\DeclareRobustCommand{\VAN}[3]{#2}
\let\VANthebibliography\thebibliography
\def\thebibliography{\DeclareRobustCommand{\VAN}[3]{##3}\VANthebibliography}

\usepackage{graphicx}	% Including figure files
\usepackage{amsmath}	% Advanced maths commands

\newcommand{\aref}[1]{\hyperref[#1]{Appendix~\ref{#1}}}

\defcitealias{KT18}{KT18}
\defcitealias{WK23}{WK23}
\defcitealias{Hu23}{H23}

\title[Galactic Metal Distributions]{Understanding the Mechanisms Behind the Distribution of Galactic Metals}

\author[Zhang et al.]{
Chuhan Zhang,$^{1}$\thanks{E-mail: chuhan.zhang@anu.edu.au (ANU)}
Zefeng Li,$^{2}$
Zipeng Hu$^{3}$
and Mark R. Krumholz$^{1,4}$
\\
$^{1}$Research School of Astronomy and Astrophysics, Australian National University, Cotter Road, Weston ACT 2611, Australia\\
$^{2}$Centre for Extragalactic Astronomy, Department of Physics, Durham University, South Road, Durham DH1 3LE, United Kingdom\\
$^{3}$Kavli Institute for Astronomy and Astrophysics, Peking University, Beijing 100871, China\\
$^{4}$ARC Centre of Excellence for Astronomy in Three Dimensions (ASTRO 3D), Canberra ACT 2601, Australia
}

\date{Accepted XXX. Received YYY; in original form ZZZ}

\pubyear{\the\year{}}

\begin{document}
\label{firstpage}
\pagerange{\pageref{firstpage}--\pageref{lastpage}}
\maketitle

\begin{abstract}
The evolution and distribution of metals within galaxies are critical for understanding galactic evolution and star formation processes, but the mechanisms responsible for shaping this distribution remain uncertain. In this study we carry out high-resolution simulations of an isolated Milky Way-like galaxy, including a star-by-star treatment of both feedback and element injection. We include seven key isotopes of observational and physical interest, and which are distributed across different nucleosynthetic channels—primarily AGB stars (N, Ba, Ce), supernovae (O, Mg, S), and Wolf-Rayet stars (C)show measurably diﬀerent correlation statistics in space and time and their ﬂuctuations. This diﬀerence arises from the distinct ejection mechanisms associated with each nucleosynthetic process. The large-scale properties ensure that diﬀerent elements, despite having diﬀerent nucleosynthetic origins, are highly correlated with one another (> 0.85 for all, > 0.99 for same origins), and their spatial correlations vary together in time. However small-scale variations naturally break elements into distinct nucleosynthetic familiars, with elements originating from the same channels correlating better with each other than with elements from different origins. Our findings suggest both challenges and opportunities for ongoing efforts to use chemical measurements of gas and stars to unravel the history and physics of galaxy assembly.
\end{abstract}

%%%%%%%%%%%%%%%%%%%%%%%%%%%%%%%%%%%%%%%%%%%%%%%%%%

%%%%%%%%%%%%%%%%% BODY OF PAPER %%%%%%%%%%%%%%%%%%

\section{Introduction}

% draw out the importance
The distribution and evolution of metals (elements heavier than helium) in galaxies is a key focus for understanding how galaxies form and evolve. Metals are produced primarily through stellar nucleosynthesis and redistributed into the interstellar medium (ISM) through stellar winds, supernovae (SNe), and other feedback processes. Once in the ISM, these metals are mixed by transportation processes including radial migration of metal-enriched gas across the galaxy, as well as turbulence, affecting future star formation and drives the chemical evolution of galaxies (for reviews, see \citealt{Tinsley_1980, Maiolino_2019, Sanchez_2021}). 

% introduce observational background
Metallicity, the oxygen abundance in ionised regions, can be measured using emission line diagnostics (for a review, see \citealt{Kewley_2019}). The measurements of metallcities have been extended from integrated single fibre technique \citep[e.g.][]{Tremonti04} to spatially resolved technique (e.g. \citealt{2011A&A...534A...8M, 2012MNRAS.421..872C, 2012A&A...538A...8S, 2015ApJ...798....7B, 2019MNRAS.484.5009E, 2020AJ....159..167L, 2022A&A...659A.191E}). The deployment of integral field units (IFUs) enables  measurements of the spatially resolved two-dimensional distributions of oxygen abundance across nearby galaxies (e.g. \citealt{2011MNRAS.415.2439R, 2016ApJ...830L..40S}). These observations reveal the existence of metallicity gradients, typically showing that metal abundance decreases from the center of galaxies outward (e.g. \citealt{2017MNRAS.469..151B, 2018A&A...618A..64H, 2018MNRAS.479.5235P, 2018A&A...609A.119S, 2019ApJ...887...80K}). 
% introduce how people study the mechanisms by gradients
A range of theoretical studies have aimed to explain the origin of these gradients and situate them in the broader context of galaxy formation (e.g. \citealt{2009A&A...499..427D, 2017MNRAS.466.4780M, 2021MNRAS.502.5935S, 2022MNRAS.511.1667T}).

% introduce 2-point correlation funcitons; analytical model KT18 based on 2-point correlation functions; current work of 2-point and KT18.
However, gradients represent a significant simplification of the data, since they collapse complex two-dimensional (2D) maps down to a single linear fit. To exploit the full power of IFU metallicity maps, higher-order statistics are in need to decode the detailed information of the data, which in turn can advance our understanding of how metals are injected and mixed in galaxies. One of the simplest statistical descriptions for a 2D map is the two-point correlation function, which describes the characteristic size scales over which maps vary. \citet[hereafter \citetalias{KT18}]{KT18}, provide a minimal theoretical model to predict two-point correlations of galaxy metallicities based on the competition between metal injection and diffusion processes. This prediction has motivated various observational studies to examine two-point correlations (or similar statistics), in different samples of nearby galaxies (e.g. \citealt{2020MNRAS.499..193K, 2021MNRAS.504.5496L, 2023MNRAS.518..286L, 2021MNRAS.508..489M, 2022MNRAS.509.1303W, 2024arXiv240704252L}). All these studies find that metallicity maps of nearby galaxies contain statistically-significant spatial structure on top of the overall gradient, and that the two-point correlation functions describing this structure generally follow the shape predicted by \citetalias{KT18}. These studies indicate that nearby galaxy metallicity distributions are corelated on characteristic scales of $\sim$ kpc, but with significant systematic variations with galaxy properties such as stellar mass and star formation rate. 

While this analytic and observational work has begun the study of metallicity distribution statistics, there have been few simulation efforts to date. The only published study thus far to focus on two-point statistics instead of just gradients is from \cite{2024MNRAS.528.7103L}, who post-process the Auriga cosmological simulations \citep{2017MNRAS.467..179G} to produce metallicity maps comparable to those accessible via observations. They find that the simulations successfully reproduce the correlation lengths observed in local galaxies, suggesting that they capture the most important processes associated with metal mixing in the ISM. The metal distributions produced in the simulations are also in reasonably good agreement with the predictions of \citetalias{KT18}.

% The limitation of current studies: 1. limitation of observations and simulations; 2. limitation of KT18 model.
All studies to date, however, have significant limitations. The observations generally have at best resolutions of a few hundred pc -- the main exception is studies of very nearby galaxies with PHANGS-MUSE, which resolution reaches $\sim 50$ pc, but only for samples of $\lesssim 10$ galaxies \citep[e.g.,][]{2020MNRAS.499..193K, 2022MNRAS.509.1303W}. Moreover, the studies published to date exclusively focus on oxygen, which is easier to measure in the gas phase than other elements given the sensitivity and spectral coverage of current IFUs. This will only begin to change for large samples once BlueMUSE comes online \citep{BlueMUSE} and provides access to key diagnostic lines for nitrogen.

The main limitation for simulations to date is also resolution. While the Auriga simulations \citep{2017MNRAS.467..179G} seem to capture the rough outlines of metal mixing, their effective resolution of $\sim 100$ pc may be adequate to resolve mixing driven by large-scale mechanisms such as bars \citep{2013A&A...553A.102D}, gravitational instabilities \citep{2015MNRAS.449.2588P}, cosmological accretion \citep{2016MNRAS.457.2605C}, and supernova-driven turbulence \citep{2002ApJ...581.1047D, 2017MNRAS.467.2421C}, but is clearly insufficient to resolve some other possibly-important processes such as spiralarm  \citep{2016MNRAS.460L..94G, 2023MNRAS.521.3708O} and thermal instability-driven mixing \citep{Yang_2012}. Indeed, as \citetalias{KT18} point out, in a modern spiral the mean ISM density is $\sim 1$ hydrogen atom per cm$^{3}$ and the typical gas scale height is $\sim 100$ pc, so simulations with mass resolutions of a few thousand M$_\odot$ -- typical of even zoom-in cosmological simulations -- resolve the scale height of the ISM by only $\sim 1-2$ fluid particles. Even marginal resolution of the internal structure of the ISM requires mass resolutions $\sim 100$ M$_\odot$, or spatial resolutions $\sim 10$ pc for Eulerian codes. To date no cosmological simulations of large spiral galaxies satisfy this requirement.

A further limitation of both current theory and observations is that they tell us little about the spatial relationships between different elements. As noted above, due to the limited wavelength coverage of currently available high spatial resolution IFUs,  observations of gas-phase elements to date focus almost exclusively on oxygen. By contrast, however, a much larger set of elements are available in stellar spectra. These measurements do not directly provide the spatial distribution of metals in the ISM, because as stars migrate after their formation their records of elemental abundances de-correlate with their current location in space. However, any spatial correlations between elements that were present at the time of star formation remain frozen into the distribution of stars in chemical abundance space, which persists for long times. There have been extensive efforts to understand the structure of this chemical space, since it matters for a wide variety of studies that rely on stellar abundances \citep[e.g.,][]{Bland-Hawthorn10a, Bland-Hawthorn16a, Krumholz19a, Weinberg19a, Weinberg22a, Ting22a}. In recent years, however, hese efforts, however, have for the most part focused on empirical attempts to deduce the structure of chemical space from stellar spectra. There are no first-principles predictions, for example, about how well different elements correlate, and almost none (beyond some general arguments in \citetalias{KT18}) about how the spatial statistics of elements differ depending on their nucleosynthetic origin. 

% our motivation/aims.
These situations motivate us to perform high-resolution simulations with both (1) the ability to capture the detailed structure at or better than the best-resolved current gas-phase metallicity maps, in anticipation of future higher-resolution facilities capable of capturing even more detailed structure, and (2) the ability to track multiple elements and study the relationships between them. We focus on an isolated Milky-Way-like galaxy at $z = 0$, following the production of multiple elements injected using a star-by-star treatment of feedback and nucleosynthesis, and tracing the subsequent transport of those elements through the ISM. We carry out these simulations until the statistics of their fluctuation distributions reach steady states, and then use the resulting steady-state 2D metal fluctuation maps to study the spatial statistics of multiple elements.

% structure
The outline of this paper is as follows. In \autoref{sec:sim}, we describe our isolated galaxy simulation, including the numerical methods and initial conditions, along with the statistical tools we use to analyse the metallicity distributions in space and time. We describe the outcome of our simulations and the statistical properties of the metal field that we derive from them in \autoref{sec:result}. In \autoref{sec:discussion} we discuss the results, drawing several conclusions about both the simulations and existing models for metallicity distributions. Finally, we draw conclusions in \autoref{sec:conclusion}.

\section{Simulation}\label{sec:sim}

In this paper we simulate an isolated, magnetised Milky Way-like disc galaxy. To produce a realistic gaseous metal distribution, we follow the return of metals from every single star individually, following the injection of metals back into the surrounding gas as a part of stellar feedback. Once they are injected we treat the metals as passive scalars. In the remainder of this section we describe our simulation methods, initial conditions, and statistical analysis techniques. 

\subsection{Numerical method}

Our simulation is an extension of the full galaxy zoom-in simulations described by \citet[hereafter \citetalias{WK23}]{WK23} and \citet[hereafter \citetalias{Hu23}]{Hu23}, and with the exception of some aspects of the treatment of star formation, feedback, and metals, our numerical methods are identical to theirs. For this reason we simply summarise the parts of our method that are the same here, referring readers to those papers for full details, and focus most of our attention on the modifications we have made to trace metals.

Our simulations use the \textsc{gizmo} code \citep{2015MNRAS.450...53H}. We use \textsc{gizmo}'s meshless finite mass (MFM) method for MHD, and we implement gas cooling using the GRACKLE library \citep{2017MNRAS.466.2217S}; as discussed in \citetalias{WK23}, \textsc{gizmo}'s default implementation of cooling does not correctly produce a multiphase interstellar medium, and this is not suitable for a simulation such as ours that resolves the phase structure of the ISM. We also enable \textsc{gizmo}'s subgrid turbulent edding mixing model \citep{2017MNRAS.467.2421C,2018MNRAS.480..800H}.

Our simulation uses a customised treatment of star formation and stellar feedback. Our treatment of star formation is that if the density of a gas particle exceeds a critical density $\rho_\text{c}$, we assign it a local volumetric star formation rate $\dot{\rho}_\text{SFR} = \epsilon_\text{ff}\rho_g/t_{\text{ff}}$, where $\rho_g$ is the gas particle density, $\epsilon_\text{ff}$ is the star formation efﬁciency parameter, and $t_{\text{ff}} = \sqrt{{3\pi}/{32G\rho_g}}$ is the local gas free-fall time. We adopt $\epsilon_\text{ff} \approx 0.01$ as shown by a wide range of observations \citep{Krumholz19a}, and set the critical density to $\rho_\text{c} = 10^3~\text{H}~\text{cm}^{-3}$, which given our mass resolution and cooling curve is roughly equivalent to setting $\rho_\text{c}$ such that Jeans mass is equal to the mass resolution for particles with density $\rho_\mathrm{c}$ and temperatures at the equilibrium value for that density -- see \citetalias{WK23} for details. We similarly adopt a minimum softening length of gas particles of 0.1 pc, which is roughly the Jeans length at $\rho_\mathrm{c}$ and the equilibrium temperature. To avoid spending too much computational time following very dense structures, for particles with $\rho_g > 10^2 \rho_\text{c}$ we set $\epsilon_\text{ff} = 10^6$ so that they are converted to stars nearly instantaneously. Thus our composite expression for the volumetric star formation rate is
\begin{equation}
\dot{\rho}_\text{SFR} = 
\begin{cases} 
0 & \rho_g < \rho_{\text{c}}, \\
\epsilon_\text{ff}\rho_g/t_{\text{ff}} & \rho_{\text{c}} \leq \rho_g < 10^2\rho_{\text{c}} ,  \\
10^6\rho_g /t_{\text{ff}} & \rho_g \geq 10^2\rho_{\text{c}},  \\
\end{cases}
\label{eq:sfr}
\end{equation}
As usual, we implement this probabilistically, so that during a time step of size $\Delta t$, a particle of density $\rho_g$ has a probability $P = 1 - \exp(-\dot{\rho}_\mathrm{SFR}\Delta t/\rho_g)$ of being converted to a star particle.

Once star particles form, we carry out star-by-star tracking of feedback and metal injection; this is in contrast to the default \textsc{gizmo} treatment of an IMF-integrated stellar population, which is not suitable for the resolutions we reach where the expected number of SNe per star particle is $\sim 1$. In our simulation, when a star particle forms we draw a synthetic stellar population for that star particle using the \textsc{slug} stochastic stellar population synthesis code \citep{2012ApJ...745..145D,2015MNRAS.452.1447K}. The stars are drawn from a Chabrier (2005) IMF \citep{Chabrier05a}, using a fully stochastic treatment for all stars above $1$ M$_\odot$ in mass. Each star follows an individual evolutionary track, the Padova stellar tracks \citep{2012MNRAS.427..127B}, and at each time step we model its atmosphere using \textsc{slug}'s ``starburst99'' option for stellar atmosphere models \citep{1999ApJS..123....3L}. This allows us to calculate the instantaneous ionising luminosity of each individual star particle, taking into account the unique properties of the stars that contribute to the population, which we inject back into the simulation domain using a Str\"omgren volume method to calculate the effects of ionisation feedback. Similarly, we track which stars end their lives as supernovae, injecting mass, energy, and metals and which end their lives as asymptotic giant branch (AGB) stars, injecting mass and metals but no energy.

Our treatment of SN energy injection follows the approach described in \citet{2019MNRAS.490.4401A}, which is a variant of the common approach of injecting $10^{51}$ erg of energy in regions where the density is low enough that the Sedov-Taylor phase of SN blast wave expansion can be resolved, and gradually changing over to injecting radially-outward momentum as the resolution degrades; see \cite{2019MNRAS.490.4401A} for details. 

With regard to mass and metal return, we track the distribution of several key isotopes, treating each one as a passive scalar. The isotopes we include are $^{12}$C, $^{14}$N, $^{16}$O, $^{32}$S, ${^24}$Mg, $^{138}$Ba, and $^{140}$Ce. Among these, $^{12}$C, $^{14}$N, $^{16}$O, and $^{32}$S are dominant in the interstellar medium, while $^{24}$Mg, $^{138}$Ba, and $^{140}$Ce are crucial in tracing processes within stars. Initially, all isotope abundances are set to zero across the simulation. As the simulation progresses, isotopes are injected into the gas as part of the stellar feedback process. Specifically, in each time step for each star particle, we calculate the real-time yields of each isotope released by stars based on their mass and evolutionary stage. To determine these yields, we rely on three different yield tables: \citet{2016ApJ...821...38S} for type II supernovae, \citet{2016ApJ...825...26K} for AGBs, and \citet{2014MNRAS.437..195D} for super-AGBs. After obtaining the isotope yields for each stellar particle, we enrich the surrounding gas particles. The newly released total mass and isotopes are distributed to the gas particles around the star following the same algorithm that is used for distributing supernova energy distribution \citep{2018MNRAS.480..800H}. In essence, the mass is distributed in proportion to the overlap between the stellar particle's position and the neighboring gas particles's weighting kernels.

\subsection{Initial conditions}
\label{ssec:initcon}

The initial condition for our simulation is a snapshot taken from the simulation of an isolated Milky Way-analog galaxy by \citetalias{WK23}, which was also performed using \textsc{gizmo} and including a very similar treatment of star formation and feedback to the one we adopt here, but with lower resolution, an IMF-integrated rather than a star-by-star treatment of feedback to accord with this lower resolution, and no tracking of metals. To generate an initial condition suitable for re-simulation at higher resolution, we follow the procedure outlined in \citetalias{Hu23}.

We start from the 600 Myr snapshot of \citetalias{WK23}, which has a gas fraction similar to the present-day Milky Way, and we switch from the IMF-integrated treatment of feedback in that simulation to the star-by-star treatment described above. We then advance the simulation for 100 Myr, during which time we continue to disable metals and we leave the resolution unchanged; our goal during this phase is to build up a realistic population of stellar particles that can provide feedback and metals. Because the integrated amount of feedback is the same for our IMF-averaged and star-by-star approaches, the star formation rate and other properties of the simulation remain stable during this time. After this 100 Myr interval, we subdivide particles in the simulation to increase the resolution from the original 859.3 $\text{M}_\odot$ used in \citetalias{WK23} to 286.4 $\text{M}_\odot$, following the particle splitting method described by \citetalias{Hu23}; we also turn on metal injection and diffusion at this point. We will refer to the state that exists at this point as time $t=0$ from this point on. Due to the sudden increase in mass resolution that occurs at this time, the star formation rate (SFR) in the simulation undergoes an initial fluctuation, but then settles back into a steady state at nearly the same SFR as before we increased the resolution within $\lesssim 100$ Myr. We use the properties of the simulation after this point as the basis for our statistical analysis below.

\begin{figure*}
	\includegraphics[width=\textwidth]{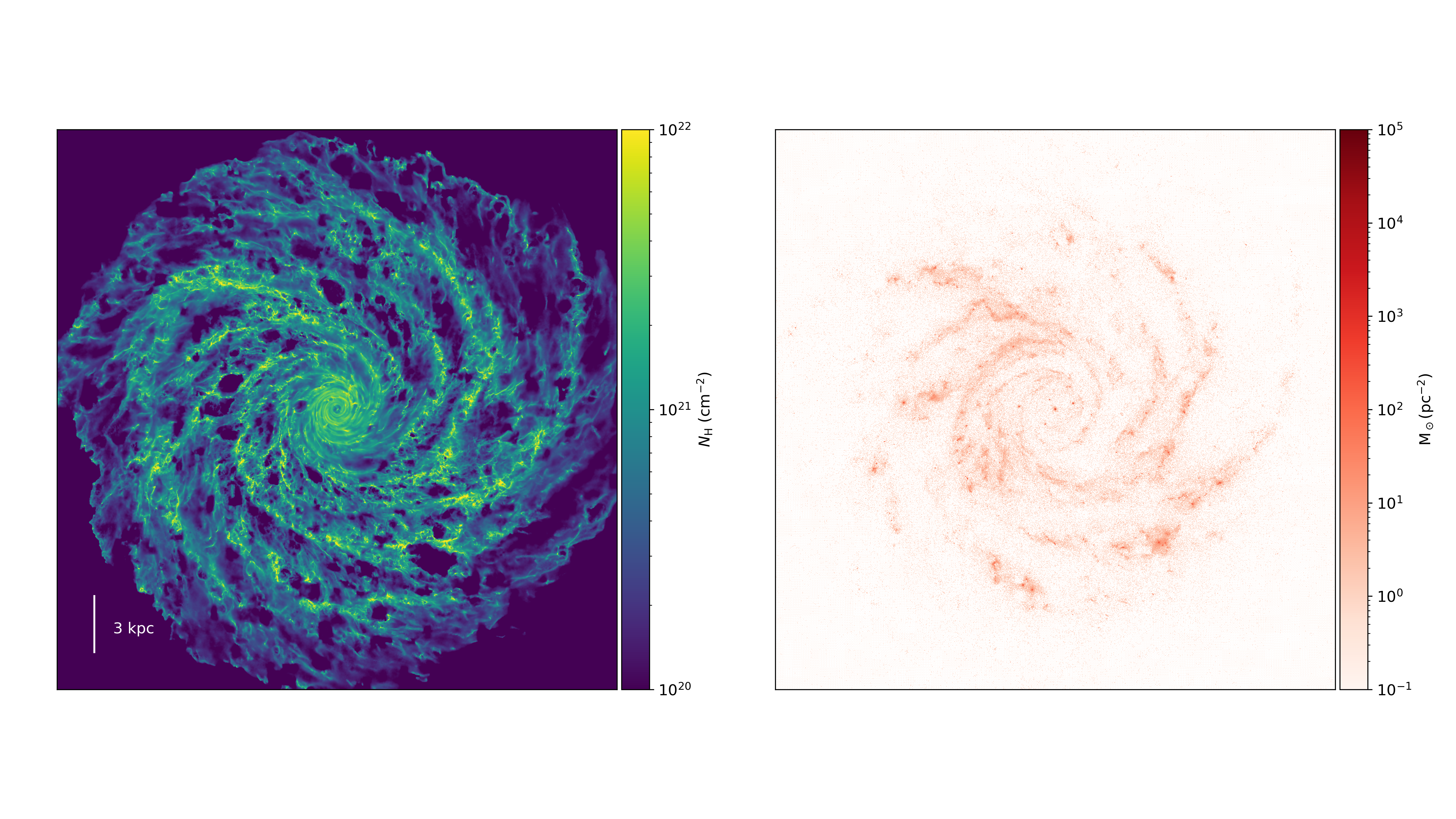}
        \caption{Snapshot of the simulation at $t=5$ Myr. The panels from left to right shows the total surface density of the gas (expressed as number of H nuclei per unit area) and stars respectively. The dimensions of each map are 30 $\times$ 30 kpc, while the pixel size of each map is 20 pc.}
        \label{fig:ini}
\end{figure*}

Due to the splitting, the smoothing length of the gas particles needs to be adjusted by \textsc{gizmo} in the initial few Myr. To illustrate the initial conditions of our simulation, we demonstrate the snapshots at 5 Myr in \autoref{fig:ini} when the smoothing length has been fixed.

\subsection{Analysis methods}\label{sec:tool}

The central object of our study is the statistics of metallicity fluctuations in galaxies. Here we explain the steps in the analysis pipelines we use to derive these statistics from the raw simulation outputs; unless otherwise stated, we carry out these steps for each time snapshot.

\subsubsection{Characterising the metal distribution}
\label{sec:metalfields}

The first step in our analysis chain is to convert the three-dimensional simulated metal distribution into a face-on projection of the metallicity $Z$ for each element as a function of position in the plane of the galactic disc. To this end, we use \textsc{yt} \citep{2011ApJS..192....9T} plus \textsc{Meshoid} \citep{Grudic_meshoid_2021} to calculate the surface density of total gas mass and the mass of each element projected along the $z$-direction, which is defined as the direction orthogonal to the disc plane; formally, we write these surface densities as $\Sigma = \int \rho \, dz$ and $\Sigma_Z = \int \rho_Z \, dz$, where $\Sigma$ is the total gas surface density, $\Sigma_Z$ is the surface density of some element, and both of these quantities are functions of the position $(x,y)$ in the galactic plane. We carry out this calculation on a 2D grid of size $30\, \mathrm{kpc}\times 30\,\mathrm{kpc}$ with spacing $\Delta x = \Delta y = 37.5$ pc (corresponding to a resolution of $800\times 800$), centred on the centre of the galaxy. We then define the abundance field for each element by $Z = \Sigma_Z /\Sigma$.

Because the primary variation of the metal field is simply a radial gradient caused by the higher star formation rate in the galactic centre, when characterising the statistics of metal fields it is common to subtract off this gradient \citep{2019ApJ...887...80K,2020MNRAS.499..193K,2021MNRAS.504.5496L,2023MNRAS.518..286L,2024MNRAS.528.7103L}. We therefore also compute an azimuthally-averaged metallicity (again for each element) $\overline{Z}_r$ in annular bins of width $\Delta x$ centred on the galactic centre, and define a metallicity fluctuation map $Z'=Z-\overline{Z}_r$ with the gradient removed.

%In Fig. ~\ref{Fig:ini}, we show the face-on gas surface density and surface metallicity fluctuation map at 564 Myr.

%!!! Here put our final snapshot. Or we can show it later

\subsubsection{Auto-, cross-, and time-correlation}
\label{ssec:correlations}

The next step in our algorithm is to calculate the two-point auto- and cross-correlations of the metal fields. Formally, we  define these as an expanded form of \citet{2021MNRAS.504.5496L}
\begin{equation}
    \xi_{a,b}(r) = \left\langle\frac{\left\langle Z_a'(\mathbf{r} + \mathbf{r'}) Z_b'(\mathbf{r'}) \right\rangle_{\mathbf{r}'}}{\sqrt{\left\langle Z_a'^2(\mathbf{r'}) \right\rangle_{\mathbf{r}'} \left\langle Z_b'^2(\mathbf{r'}) \right\rangle_{\mathbf{r}'}}}\right\rangle_{\theta}
    \label{eq:cross_corr}
\end{equation}
where $Z_a'$ and $Z_b'$ are the metallicity fields for two isotopes $a$ and $b$; here the angle brackets indicate averages, with $\langle\cdot\rangle_{\mathbf{r}'}$ indicating an average over the dummy position variable $\mathbf{r}'$ and $\langle\cdot\rangle_\theta$ an average over the angle $\theta$ of the lag vector $\mathbf{r}$. If $a$ and $b$ are the same isotopes this represents the auto-correlation, while if they are different it represents the cross-correlation. Note that the autocorrelation is exactly equal to unity at lag $r=0$.

In practice we compute the auto- and cross-correlation by placing all pixel pairs in our maps into bins of lag, and averaging over bins \citep{2021MNRAS.504.5496L}.
Let $r_n$ be the central lag of the $n$th bin, and $r_{n-1/2}$ and $r_{n+1/2}$ be the minimum and maximum lag for that bin. We then compute the correlation function at lag $r_n$ for metal fields $a$ and $b$ as
\begin{equation}
    \xi_{a,b}(r_n) = \left(\frac{1}{\sigma_a \sigma_b}\right)\frac{1}{N_n} 
    \sum_{r_{n-1/2} < r_{ij} < r_{n+1/2}} Z'_a(\mathbf{r}_i) Z'_b(\mathbf{r}_j),
    \label{eq:corr_implementation}
\end{equation}
where $\mathbf{r}_i$ and $\mathbf{r}_j$ are the positions of pixels $i$ and $j$, $r_{ij} = |\mathbf{r}_i - \mathbf{r}_j|$ is the distance between them, the sum runs over all $N_n$ pixel pairs $(i,j)$ for which $r_{n-1/2} < r_{ij} < r_{n+1/2}$, and $\sigma^2_{a,b} = \langle Z_{a,b}'^2\rangle$ are the variances of fields $a$ and $b$ across the whole map.

\begin{figure}
    \includegraphics[width=\columnwidth]{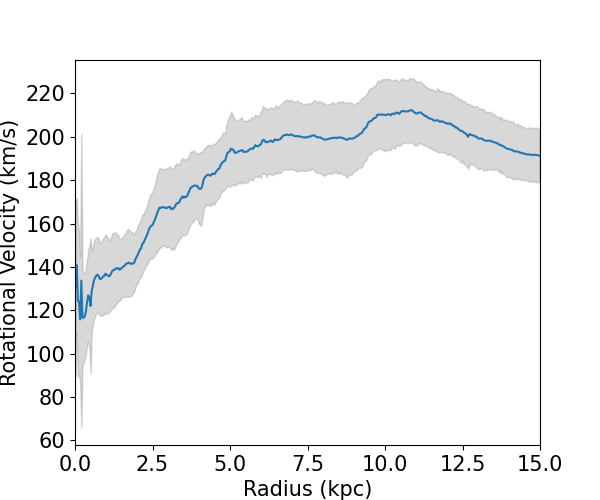}
    \caption{Galactic rotation curve computed as described in \autoref{ssec:correlations}. The blue curve indicates the mean rotational velocity as a function of galactocentric radius from 0 to 15 kpc at 564 Myr, while the grey area shows the $1\sigma$ scatter from particle to particle.}
    \label{fig:RC}
\end{figure}

In addition to the static galaxy metal fields that are accessible through observations, we can also study metal fluctuations over time in our simulations. We would therefore like to compute the two-point correlation function between two snapshots at different times. The main difficulty with this is that, just as the spatial correlation is dominated by the metallicity gradient, which we must remove if we are to study higher-order statistics, the temporal correlation will be dominated simply by the overall rotation of the galaxy. To remove this effect, we trace all gas particles back through time following the overall galactic rotation curve. We divide the distance $r$ from 0 to 15 kpc into 400 bins, so the bin width matches the sampling size in our metallicity maps. We compute the rotation curve $v_\phi$ at each galactocentric radius as the mass-weighted mean of the circular velocities of the gas particles in that radial bin, $v_\phi = {\sum m_i v_{\phi,i}}/{\sum m_i}$, where the sum runs over all gas particles $i$ in a radial bin, $m_i$ is the particle mass, and $v_{\phi,i}$ is the $\phi$ component of velocity in a polar $(r,\phi)$ coordinate system centred on the galactic centre. We show the rotation curve we derive from one of our snapshots (at $t=564$ Myr -- see \autoref{sec:result}) in \autoref{fig:RC}, and we find that there is very little variation over time, so we adopt the rotation curve shown in this figure for all times.

Having established the rotation curve, we are now in a position to define a correlation in time. While in principle we could compute the correlation as a function of both time and space lag (see \citetalias{KT18}), and between different elements, in this work we will limit ourselves to lags in time and for a single element only. We therefore define the time correlation as
\begin{equation}
        \xi(t) = \frac{\left\langle Z'(\mathbf{r}, t') Z'(\mathbf{r} - v_\phi t \boldsymbol{\hat{\phi}}, t'+t) \right\rangle_{\mathbf{r},t'}}{\sqrt{\left\langle Z'^2(\mathbf{r}, t') \right\rangle_{\mathbf{r},t'} \left\langle Z'^2(\mathbf{r}, t'+t) \right\rangle_{\mathbf{r},t'}}}.
    \label{eq:time_corr}
\end{equation}
Here $Z'(\mathbf{r},t)$ is the projected metal fluctuation field (for some element) at time $t$ and position $\mathbf{r}$, $\boldsymbol{\hat{\phi}}$ is a unit vector in the $\phi$ direction, and the notation $\langle\cdot\rangle_{\mathbf{r},t'}$ indicates an average over all positions $\mathbf{r}$ and over the dummy time variable $t'$. The term in the numerator $v_\phi t \boldsymbol{\hat{\phi}}$, where $v_\phi$ is the rotation velocity evaluated at the particle's radius, removes the effects of overall rotation of the galaxy.

In practice we implement evaluation of \autoref{eq:time_corr} on a collection of simulation snapshots by selecting pairs of outputs separated by a fixed time interval $t$.\footnote{Since the outputs are uniformly spaced in time, there is no need for binning.} We then rotate the particle positions in the later snapshot in each pair by an angle $\Delta \phi = -t (v_\phi/r)$ about the axis defined by the galactic plane and galactic centre, generate metal fields from the rotated positions as described in \autoref{sec:metalfields}, and then evaluate the two-point correlation of the two fields using \autoref{eq:corr_implementation} for a spatial lag of zero. This yields a value of the correlation for each snapshot pair with time lag $t$, and to obtain the final value of $\xi(t)$ we simply take the average over all pairs. We limit our analysis to lags of no more than 50 Myr, since we expect the de-rotation procedure to become increasingly inaccurate over longer times. Moreover, because rotating the particle positions and generating the metallicity fluctuation maps from those rotated positions is somewhat computationally intense, and we output snapshots at relatively high cadence and thus the number of possible pairs is very large, we do not use all possible snapshot pairs when evaluating \autoref{eq:time_corr}. Instead, we only perform the rotation operation on snapshots at intervals of 10 Myr, though we then compare these rotated snapshots to earlier snapshots at finer time cadences (i.e., so for example we only rotate snapshots at 100 Myr, 110 Myr, and so forth, but we still compute a correlation at a lag of 1 Myr by comparing the rotated 100 Myr and 99 Myr snapshots, the 110 and 109 Myr snapshots, and so on). We have verified that all the results we present do not change qualitatively if we alter the cadence of 10 Myr.

%\begin{figure}
%    \includegraphics[width=\columnwidth]{O16_526_6.png}
%    \caption{The left panel is the O16 fluctuation map at 520 Myr, and the right panel is counter-rotated 6 Myr from 526 Myr, of which the co-rotation is removed and the injection and diffusion remained. Therefore the time-correlation by computing with these two fluctuation maps gives us how the metal correlation decrease due to diffusion by time.}
%    \label{fig:re-rotation}
%\end{figure}

\subsubsection{Parametric parameter estimates from the correlation functions}\label{sec:model}

%The model presented in KT18 paper offers a theoretical framework for understanding the spatial distribution of metallicity in galaxies, which is based on a stochastic diffusion process, where metals are injected into the ISM through star formation events and subsequently mixed via interstellar turbulence. This model allows for a detailed analysis of the spatial and temporal fluctuations in metallicity, and gives us a more direct way to study the physical processes than just using 2-point correlation values at some defined scales. However, we do not fully trust this model as its assumptions do not fully match our simulations. We will detail how we use and how we test the KT18 model in this section.

In addition to the raw correlation functions, it is helpful to extract some values by fitting them against a parametric model, for which purpose we use the model of \citetalias{KT18}, which has been shown to provide very good fits to both observations \citep{2021MNRAS.504.5496L,2023MNRAS.518..286L} and lower-resolution cosmological simulations \citep{2024MNRAS.528.7103L}. The model predicts that the two-point correlation function as a function of spatial lag $r$ and time lag $t$ is
\begin{equation}
\begin{split}
\xi(r, t) &= \frac{2}{\sqrt{\ln \left( 1 + \frac{2 \kappa t_*}{x_0^2} \right) \ln \left[ 1 + \frac{2 \kappa (t_* - t)}{x_0^2} \right] }} \cdot \\
&\quad \int_0^\infty e^{-(\kappa t + x_0^2) a^2}\left[ 1 - e^{-2 \kappa (t_* - t) a^2} \right]\frac{J_0(ar)}{a} \, da,
\end{split}
\label{eq:kt18}
\end{equation}
where $J_0$ is the Bessel function of order zero. The parameters appearing in this expression are the diffusion coefficient $\kappa$, the injection time scale $t_*$, and the injection width $x_0$. 
%Alternately, since the diffusion coefficient appears in this expression only via the combination $\kappa t_*$, it is convenient to replace it as a parameter with the correlation length $l_\mathrm{corr} = \sqrt{\kappa t_*}$. Note that the model only makes a prediction for the auto-correlation and time-correlation, so it does not apply to the cross-correlation.

Since the correlations we have computed are evaluated either at the same time (i.e., the time lag $t=0$) or at the same position (spatial lag $r=0$), we can specialise \autoref{eq:kt18} to these two cases. This gives a spatial correlation at zero time lag
\begin{equation}
\xi(r) = \frac{2}{\ln \left( 1 + 2 \phi^2 \right)}
\int_0^\infty e^{-l_\mathrm{corr}^2 a^2/\phi^2}\left( 1 - e^{-2 l_\mathrm{corr}^2 a^2} \right)\frac{J_0(ar)}{a} \, da,
\label{eq:kt18-space}
\end{equation}
and a temporal correlation at zero spatial lag
\begin{equation}
\xi(t) =  \frac{\ln\left(1 + 2\phi^2-\phi^2 t/t_\mathrm{corr}\right) - \ln\left(1+\phi^2 t/t_\mathrm{corr}\right)}{\sqrt{\ln \left( 1 + 2 \phi^2 \right) \ln \left( 1 + 2 \phi^2 - 2 \phi^2 t/t_\mathrm{corr} \right) }},
\label{eq:kt18-time}
\end{equation}
where we have defined $l_\mathrm{corr}=\sqrt{\kappa t_*}$ as the correlation length, $t_\mathrm{corr} = l_\mathrm{corr}^2 / \kappa$ as the correlation time, and $\phi = l_\mathrm{corr}/x_0$ as the ratio of the correlation and injection lengths. We fit the measured auto-correlation function for element for each snapshot in our simulation to the functional form given by \autoref{eq:kt18-space}, and we fit the time correlations for each element to \autoref{eq:kt18-time}.  We carry out these fits using the Python package \textsc{emcee} \citep{2013PASP..125..306F}, an implementation of an affine-invariant ensemble sampler for Markov chain Monte Carlo (MCMC). For the spatial auto-correlation we use $x_0$ and $l_\mathrm{corr}$ as our fit parameters, while for the temporal correlation we use $\phi$ and $t_\mathrm{corr}$; we adopt flat priors for all values $>0$ on all these quantities. We take the (log) likelihood function for the fit to be a $\chi^2$ form given by
\begin{equation}
\ln \mathcal{L} = 
-\frac{1}{2} \sum_{i=1}^{N} \left[ \frac{(\xi_{\text{mod}} - \xi_{\text{sim}})^2}{\sigma_{\xi,\text{sim}}^2} + \ln(\sigma_{\xi,\text{sim}}^2) \right]
\end{equation}
where the sum is over all the bins of space or time lag at which we measure the simulation correlation function, $\xi_\text{mod}$ is the \citetalias{KT18} model-predicted correlation function evaluated from \autoref{eq:kt18-space} or \autoref{eq:kt18-time}, $\xi_{\text{sim}}$ is the auto- or time-correlation measured directly from the simulations, and $\sigma_{\xi,\text{sim}} = \sqrt{\mathrm{var}(\xi_\mathrm{sim}) / N_\mathrm{pair}}$ is a standard error that we set equal to the variance $\mathrm{var}(\xi_\mathrm{sim})$ of all the $N_\mathrm{pair}$ pixel or snapshot pairs that contribute to a given bin. In the MCMC fits, we use 100 walkers and run them for 1,000 steps, discarding the first 500 steps for burn-in; we choose this interval by visually examining the flatness of the chains for all parameters. We estimate posterior PDFs from the remaining 500 time steps.

\section{Results}\label{sec:result}

We first describe the overall evolution of our simulation in \autoref{ssec:sim_summary}. We examine spatial statistics of the individual element metal fields in \autoref{ssec:space_stats} and temporal statistics in \autoref{ssec:time_stats}. We finally consider the relationships between different elements in \autoref{ssec:cross_corr}.

\subsection{Overview of simulation results}
\label{ssec:sim_summary}

\begin{figure}
	\includegraphics[width=\columnwidth]{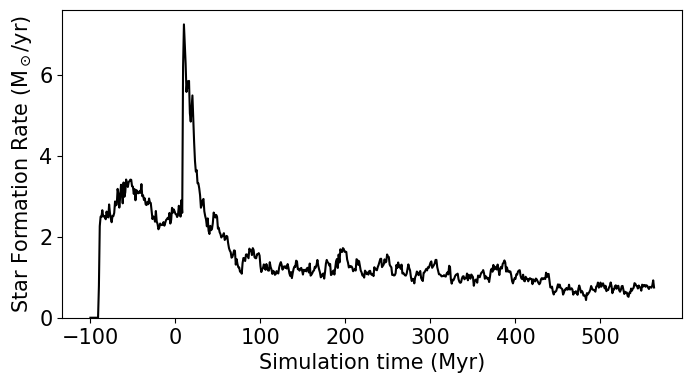}
        \caption{Star formation rate as a function of time in the simulation; time $t=0$ corresponds to the time at which we increase the resolution and turn on metal injection and diffusion, and the transient increase in star formation rate that occurs at this time is a result of the change in resolution.}
        \label{fig:sfr}
\end{figure}

We evolve the simulation from the initial condition described in \autoref{ssec:initcon} for a total of 664 Myr -- 100 Myr to $t=0$ at the original WK23 resolution with no metal injection, and then to $t=564$ Myr at increased resolution and with metals enabled. We show the star formation rate in the simulation as a function of time in \autoref{fig:sfr}. We see that there is a brief transient when we increase the resolution, but that after $\approx 50$ Myr the simulation settles back to near steady-state, giving us $\approx 500$ Myr of steady-state evolution to analyse. There is a small secular decrease in star formation rate over this time driven by gas consumption, but this is only a tens of percent-level effect.

\begin{figure*}
	\includegraphics[width=\textwidth]{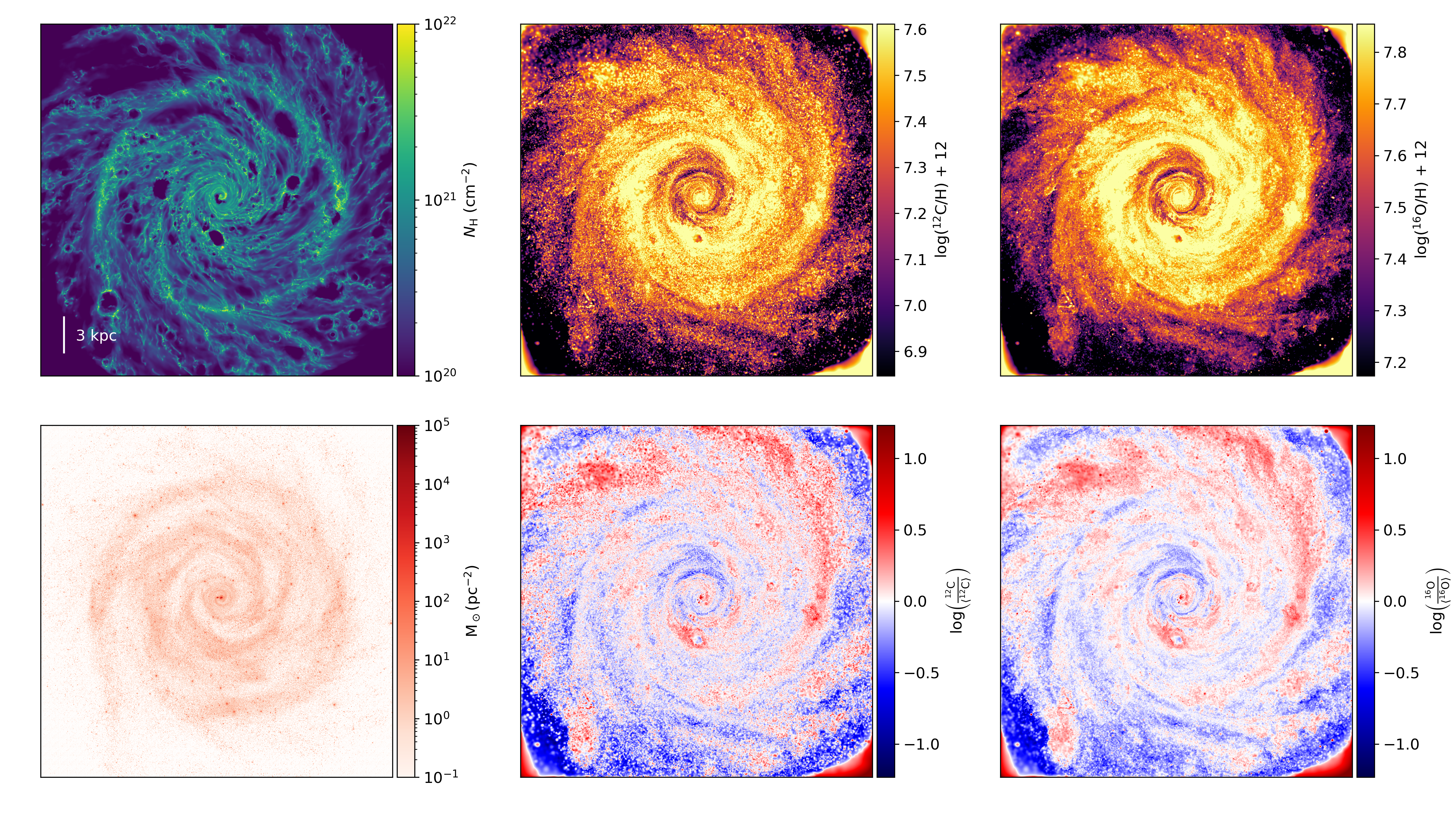}
        \caption{Snapshot of the simulation at $t=564$ Myr. The left column shows the total surface density of gas (expressed as number of H nuclei per unit area) and stars. The upper middle panel shows vertically-averaged $^{12}$C abundance $Z$, expressed in $12 + \log(^{12}\mathrm{C}/\mathrm{H})$ units, while the lower middle panel shows the abundance fluctuation map left after subtracting off the mean metallicity in annular bins (see \autoref{sec:metalfields} for details). The right column is the same as the middle column, but for $^{16}$O. The dimensions of each map are 30 $\times$ 30 kpc, while the pixel size of each map is 20 pc.}
        \label{fig:final}
\end{figure*}

We show the state of the simulation at the final time in \autoref{fig:final}. We caution that the absolute metallicities are substantially lower, and the amplitude of metallicity residuals slightly larger, than would typically be expected of spiral galaxy in the present-day Universe. This is simply the result of our having started our simulations from zero metallicity and having run them for only $0.66$ Gyr: the simulations have simply not formed enough stars over this time to match the total metallicites produced over 13 Gyr of cosmological star formation. However, this does not matter for our purpose in studying the statistics of metallicity fluctuations, because adding a constant or even a radially-varying (but azimuthally-symmetric) metallicity would not affect the statistics of interest to us (c.f.~\autoref{sec:tool}).

We can see in that the total metallicity and metallicity fluctuations partly mirror the spiral structure visible in the gas (and to a lesser extent stellar) surface density maps, but that the correspondence is not perfect. For example, there are several low-density bubbles blown driven by supernovae in the gas surface density map that show high abundances in the total metallicity and fluctuation maps. We also see that the metallicity maps for different elements are quite similar in their overall appearance, a point to which we shall return in \autoref{ssec:cross_corr}.

%It is notable that, in Fig.~\ref{fig:sfr} we find the $\text{SFR} = \text{d}M_\star/\text{d}t$ is not always stable, $M_\star$ is the star formation mass. The SFR is influenced by the change in mass resolution and the fast depletion of the dense gases at the beginning and lead to unbalanced yields of isotopes over time.

%In subsequent parts of the section we show the results obtained by the above methods and improve some of them additionally. But first, we need to do an additional processing to better apply the theoretical model. Since the spiral arms lead to an unanticipated oscillation in the KT18 model, we only fit within the first separation range before the 2-point correlation drops under 0. We will discuss it in Section \ref{subsec:spiral_arm}.

\subsection{Spatial correlations of elemental abundances}
\label{ssec:space_stats}

\begin{figure}
    \includegraphics[width=\columnwidth]{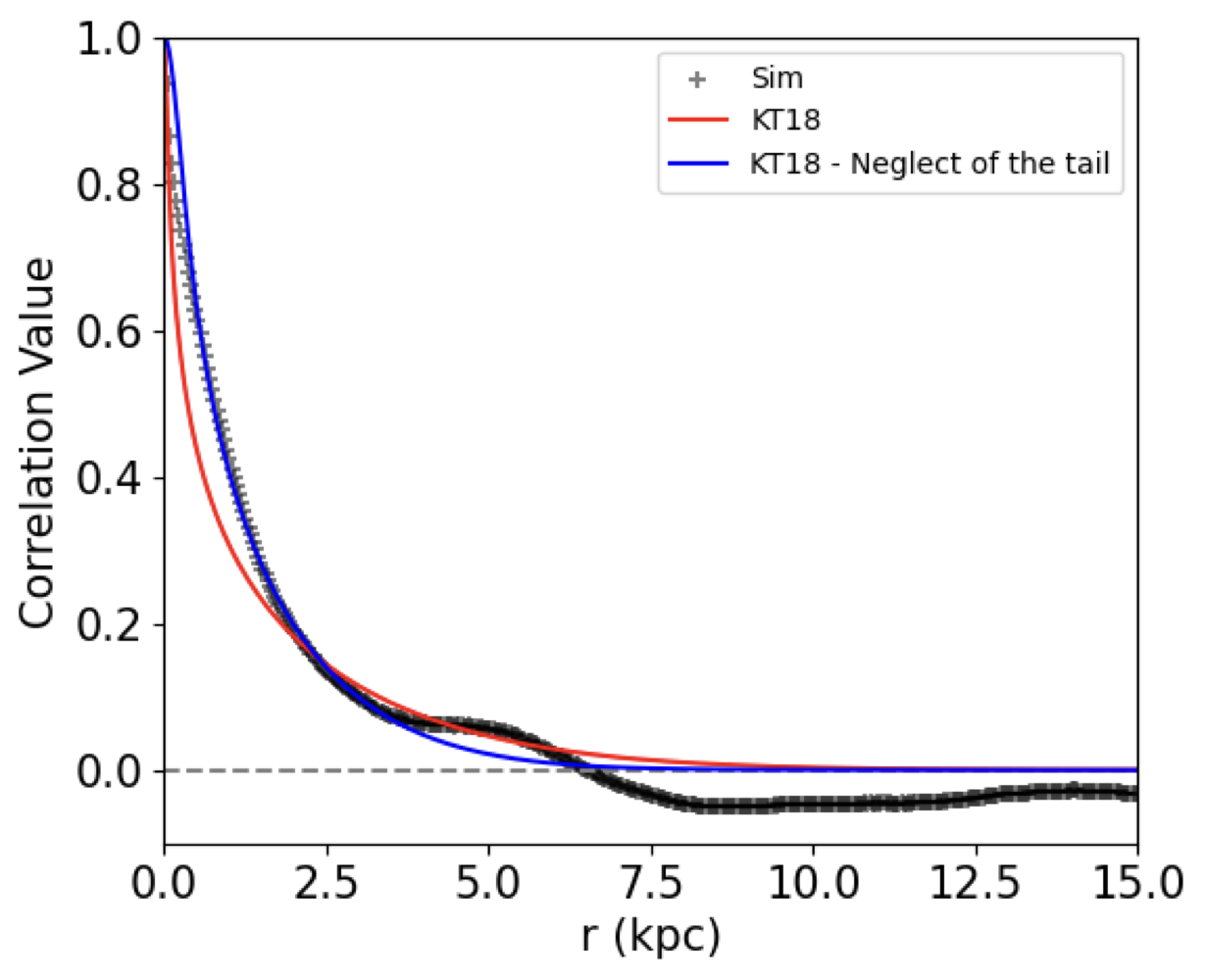}
    \caption{Auto-correlation curve of $^{14}$N at $t=450$ Myr. Black crosses indicate the values measured directly from the simulation, while the red and blue lines are fits to the \citetalias{KT18} model (\autoref{eq:kt18} in this paper) over the full 15 kpc shown and and over the range of lags smaller than the first crossing of 0.1 of the measured values at $\approx 3$ kpc, respectively; the fits shown are evaluated using the median values of each free parameter returned by the MCMC. We see that the fit over the restricted region more closely matches the data at small lags.}
    \label{fig:N14_450}
\end{figure}

We next examine the two-point auto-correlations of individual elements (\autoref{ssec:correlations}). We show an example measurement of this for $^{14}$N at the final time in our simulations in \autoref{fig:N14_450}. As the plot shows, we measure a clear auto-correlation, which is qualitatively similar to the ones seen in observed galaxies (e.g., Figure 4 of \citet{2023MNRAS.518..286L}) and in simulations (e.g., Figure 3 of \citet{2024MNRAS.528.7103L}). On the other hand, we also see clear oscillatory structure in the two-point correlation at separations of $\approx 3-8$ kpc that was not visible in earlier observational or theoretical work on metallicity autocorrelations. We return to the question of what physical mechanism is responsible for this structure in \autoref{ssec:spiral_arm}.

Given that the KT18 model does not include structures such as the one shown, and in fact the model never predicts negative values for the two-point correlation function, there is some ambiguity in how best of fit the model to the data to extract the parametric quantities discussed in \autoref{sec:model}. To handle this, we choose to fit to the KT18 model only using the two point correlation values at separations smaller than the first crossing of 0.1 of the two-point correlation function. We find that doing so yields a significantly better fit to the measured data at small lags, since the fit is no longer attempting to reproduce the negative correlations at large lags. We illustrate this in \autoref{fig:N14_450}, where we compare the outcomes of the two fitting procedures.
%is not in perfect agreement with our simulations, as Fig: \autoref{fig:N14_450} demonstrates, the correlations of nitrogen at 450 Myr drops below at some scales, which doesn't make sense in the KT18 model. To avoid this, we fit within the first separation range defore the 2-point correlation drops to 0. We will further discuss which factor that the KT18 model doesn't captured caused this feature in Section \autoref{ssec:spiral_arm}.

\begin{figure}
    \includegraphics[width=\columnwidth]{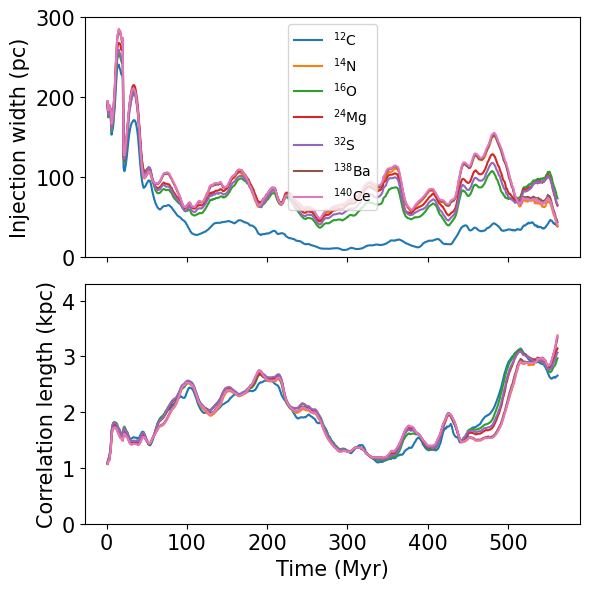}
    \caption{Best-fit injection width $x_0$ and correlation length $l_\mathrm{corr} = \sqrt{\kappa t_*}$ from parametric fits to the auto-correlations as a function of time; the values shown are the medians of the posterior PDFs returned by the MCMC fit. Colours correspond to different isotopes, as indicated in the legend.}
    \label{fig:corr}
\end{figure}

We show the median injection width and correlation length computed by our MCMC fits as a function of time in \autoref{fig:corr}; confidence intervals from the MCMC results are not shown to minimise clutter, but are generally very small, such that for most isotopes at most times they would not be visible even if we did plot them. The plot shows that correlation length of all isotopes quickly settles at a few kpc, with factor of $\sim 2$ oscillations on timescales of a few hundred Myr. We see that $^{12}$C consistently has the smallest correlation length and injection width, and that it stands out as different from all the other isotopes, which are generally clustered close to one another in both $x_0$ and $l_\mathrm{corr}$. Among the other isotopes, $^{16}$O, $^{24}$Mg, and $^{32}$S have on average slightly smaller correlation lengths and slightly larger injection widths compared to $^{14}$N, $^{138}$Ba, and $^{140}$Ce. Each of these groups of three fall nearly on top of the other members of that group. However, even between groups, oscillations of correlation length and injection width are highly-correlated, with all generally increasing or decreasing in near lockstep.
%At the end of our simulation, the correlation length of O16 and S32 reaches the magnitude of \(4\times10^3 \) pc, the correlation length of N14 is slightly shorter than them, and the correlation length of C12 ends at \(3\times10^3 \) pc. The oscillation in correlation length occurs from 200 to 500 Myr at around several kpc. %needs further study, for now we simply regard this as the result of velocity differential on the rotating arm. The differential rotation of galaxies leads to constant reorganisation of the spiral arm region, and the correlation length oscillates with the fragmentation-reorganisation of the spiral arm region. Abundance pattern resets on about 100 Myr time scales. Since the injection process always occurs centrally in the region of the spiral arm, strictly, the absolute state of equilibrium can never be reached, no matter how long it takes to diffuse, unless the galaxy has ceased its star-forming activity. However, we can pursue a not-so-strict equilibrium, where the instantaneous injection of spiral arms is no longer dominant on galaxy scales and the overall injection-diffusion forms an equilibrium. According to Fig: ~\ref{fig:O16_azi}, if we sum over azimuth from 0 to 180 degree, the correlation value is always greater than 0, which means we can roughly ignore the affect of the spiral arm at 564 Myr (discussed in Section \ref{ssec:spiral_arm}). Therefore we can roughly regard our simulation has reached the equilibrium stability at 564 Myr. The correlation curve will still oscillate but won't be far from the current values.

%It is worth warning that, some of the injection width we obtained is smaller than the sampling scale 37.5 pc, shown in the top panel as grey region, which may be unphysical. Although our resolution has improved a lot, the ability to resolve injection width is still lacking. Also, L1 norm in Fig: ~\ref{fig:corr} shows a mismatch between model and simulation. Despite our restrictions in the fitting region, other unconsidered factors still lead to biases in the results. As To be more convincing, we can just measuring the length scales at which the correlation passes through certain values to compare to fits to the KT18 model.

\begin{figure}
    \includegraphics[width=\columnwidth]{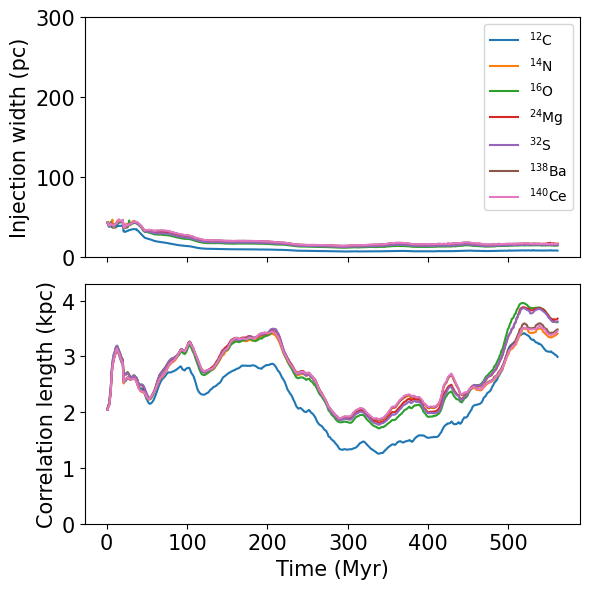}
    \caption{Same as \autoref{fig:corr}, but now showing injection widths and correlation lengths derived using a model-independent, non-parametric fitting method; see main text for details.}
    \label{fig:corr_value}
\end{figure}

Given that the KT18 model does not reproduce some of the major qualitative features of the data, one might worry about parameters extracted based on it. In order to evaluate whether this is a problem, we can also carry out a \textit{non-parametric} fit. We note that for the functional form predicted by the KT18 model at zero time lag (\autoref{eq:kt18}), the value of the correlation at 1 correlation length is always around 0.15 for physically reasonable values of the injection width, and the value of the correlation function evaluated at 1 injection width is always around 0.97 independent of the value of the correlation length. We can therefore determine model-independent estimates of the injection width and correlation length simply by measuring for what lags the measured auto-correlations pass through 0.97 and 0.15, respectively; the results we obtain by doing so should closely match the parametric values obtained by fitting to the KT18 in cases where that model is a good description of the data. In practice we implement this measurement by constructing a cubic spline interpolation of the correlation curve and measuring the lags for which this interpolated correlation curve drops to 0.97 and 0.15. We plot the results of this exercise in \autoref{fig:corr_value}.

We can see that the injection widths we obtain are significantly smaller than those that emerge from the parametric fit, which suggests a systematic issue with the \citetalias{KT18} model, a topic we will discuss further in \autoref{ssec:winj}. By contrast the correlation lengths we obtain from this model-independent fitting procedure is quite similar to those that result from the parametric fits; typical differences are tens of percent, we see the same ordering of isotopes from smallest to largest correlation length, and we see the same major qualitative shape of increase and decrease over time. We therefore conclude that our parametric correlation lengths are robust, and reflect real features present in the underlying spatial auto-correlations.

\subsection{Temporal correlations of elemental abundances}
\label{ssec:time_stats}

\begin{figure}
    \includegraphics[width=\columnwidth]{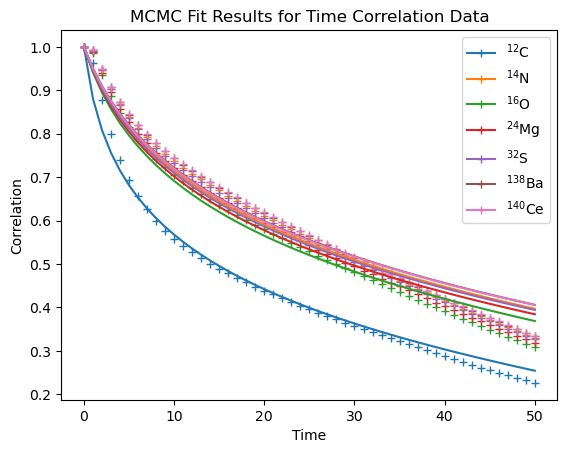}
    \caption{Crosses show the measured time correlation over 0-50 Myr lags while the centre are the mean values using snapshots from 304 - 564 Myr, following the procedure described in \autoref{ssec:correlations}. Lines show \citetalias{KT18} model fits to these data, computed using the median values of the posterior PDFs derived by fitting the average measured time correlations to the functional form given by \autoref{eq:kt18}. Colours correspond to different isotopes, as indicated in the legend.}
    \label{fig:time_corr}
\end{figure}

We next examine correlations in time, limiting our analysis to lags of $0-50$ Myr as described in \autoref{ssec:correlations}.
We evaluate the temporal correlations over this range of lags using data from 304 - 564 Myr, the time period over which our correlation lengths seem to have settled to statistical steady-state (c.f.~\autoref{fig:corr}). However, we caution that we still find a slow secular increase in temporal correlation over this period, so for example if we compute the time correlation at a fixed lag using snapshots taken from $304 - 434$ Myr versus from $434 - 564$ Myr, the latter are typically $\sim 10\%$ larger for all isotopes. This suggests that our simulation may not evolve long enough to fully converge on the time correlation.

With this caveat in mind, we show the time-correlation we compute in \autoref{fig:time_corr}. We see that elemental abundance patterns decorrelate over timescales of tens of Myr. As we saw for the correlation length, $^{12}$C stands out as the least-correlated of the isotopes we follow, while the remaining isotopes are bunched more closely together. In addition, we found that the \citetalias{KT18} model slightly overestimates the correlation time, which may be due to the differential rotation of the gas, leading to additional de-correlation beyond the effects of metal diffusion itself.

We report the results of the MCMC fit to \autoref{eq:kt18-time}
%and compare the ratios of the temporal correlation and injection lengths ($\phi_t$) and the ratio of the spatial correlation and injection lengths ($\phi_r$) 
in \autoref{tab:time}. The timescale $t_\mathrm{corr}$ for isotopes to be fully de-correlated ($\xi(t_\mathrm{corr})=0$)is smallest for $^{12}$C at $\sim 111$ Myr, while all other isotopes take about the same amount of time, $\sim 170$ Myr. In the Table we also compare the ratio of correlation length and injection width $\phi$ we derive from fitting the time correlation, which we denote $\phi_t$, with the value derived from the fits to the spatial correlation presented in \autoref{ssec:space_stats}, which we denote $\phi_r$; for the latter, the quantity we report in the table is the median and 16th to 84th percentile range of the medians at each snapshot from 304 to 564 Myr, the same as the time period over which we fit the temporal correlation. We find that the confidence intervals for $\phi_t$ and $\phi_r$ are not entirely consistent, with the values derived from the spatial correlation larger. The discrepancy is largest for $^{12}$C.

\begin{table} % Updated fit results from average
    \centering
    \caption{Result from fitting a \citetalias{KT18} model to the measured time correlations shown in \autoref{fig:time_corr} and the results of spatial correlations shown in \autoref{fig:corr} over the corresponding time range for each isotope; see \autoref{ssec:time_stats} for details on the fitting procedure. The values for both parameters of the temporal correlation (\(\phi_t\) and \(t_\mathrm{corr}\)) are the median and 68\% confidence intervals of the posterior PDF returned by the fit, while the value we report for the spatial correlation (\(\phi_r\)) is the median and 68\% range over the time 304 - 564 Myr used to fit the time correlation (c.f.~\autoref{fig:corr}).
    \label{tab:time}
    }
    \renewcommand{\arraystretch}{1.4}
    \begin{tabular}{cccc} % columns, alignment for each
        \hline
        Isotope & $\log\phi_t$ & $\log\phi_r$ & Correlation time $t_\mathrm{corr}$ [Myr] \\
        \hline
        $^{12}$C   & 1.00$^{+0.33}_{-0.25}$ & 1.86$^{+0.18}_{-0.26}$ & 111 $^{+92}_{-36}$ \\
        $^{14}$N   & 0.83$^{+0.42}_{-0.24}$ & 1.27$^{+0.30}_{-0.18}$ & 174 $^{+295}_{-68}$ \\
        $^{16}$O   & 0.84$^{+0.38}_{-0.25}$ & 1.40$^{+0.20}_{-0.16}$ & 152 $^{+202}_{-57}$ \\
        $^{24}$Mg  & 0.83$^{+0.40}_{-0.24}$ & 1.30$^{+0.26}_{-0.16}$ & 170 $^{+250}_{-67}$ \\
        $^{32}$S   & 0.85$^{+0.41}_{-0.25}$ & 1.34$^{+0.24}_{-0.16}$ & 166 $^{+265}_{-65}$ \\
        $^{138}$Ba & 0.83$^{+0.40}_{-0.25}$ & 1.26$^{+0.31}_{-0.16}$ & 178 $^{+271}_{-70}$ \\
        $^{140}$Ce & 0.84$^{+0.43}_{-0.24}$ & 1.25$^{+0.31}_{-0.17}$ & 182 $^{+323}_{-74}$ \\
        \hline
    \end{tabular}
\end{table}

\subsection{Cross-correlation between elements}
\label{ssec:cross_corr}

\begin{figure}
    \includegraphics[width=\columnwidth]{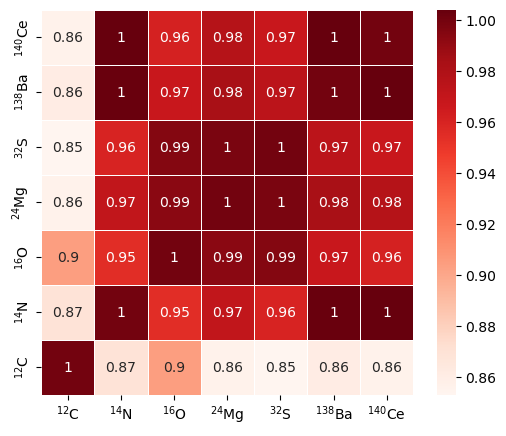}
    \caption{This map shows the cross-correlation among $^{12}$C, $^{14}$N, $^{16}$O, $^{24}$Mg, $^{32}$S, $^{138}$Ba, and $^{140}$Ce.}
    \label{fig:cross}
\end{figure}

Finally, we compute the cross correlations between all the isotopes we follow. Although we are free to compute this cross-correlation at any lag, the scientifically interesting value is the cross-correlation at zero lag, since this represents the pattern of gas-phase elemental abundances that will be frozen into newly-formed stars, and thus will shape the eventual stellar abundance distribution in chemical space. We show this quantity in \autoref{fig:cross}. Recall that a value of unity indicates perfect correlation, negative one indicates perfect anti-correlation, and zero indicates no correlation.

The first remark to make based on the Figure is that all of the elements are quite well-correlated, with even the least-correlated elements we follow showing correlations of 0.85. We remind readers that this is \textit{not} simply a result of the overall metallicity gradient in the galaxy, since we have subtracted that off and are comparing only the fluctuation maps that remain after doing so. Nonetheless, we find that to first order all of the isotopes we follow vary up and down together.

A second observation is that, although all elements are highly correlated, we can still roughly divide them into three groups, which are even better-correlated internally, and somewhat less correlated with elements from other groups. These groupings are the same ones that we have seen when examining spatial and temporal correlation statistics. One consists of $^{16}$O, $^{24}$Mg, and $^{32}$S, which form the central block in \autoref{fig:cross} and are nearly-perfectly correlated with each other (correlation coefficient 0.99). The second group consists of $^{14}$N, $^{138}$Ba, and $^{140}$Ce, which are almost perfectly correlated with each other, but slightly less correlated with the first group. Finally, we have $^{12}$C in a group of its own, showing the least correlation with the other two groups. This division of the isotopes into three groups mirrors the divisions we saw in the spatial correlation and, at least partially, in the temporal correlation. We attempt to understand the origin of this behaviour in \autoref{ssec:yield_delay}.

\section{Discussion}\label{sec:discussion}

%In \autoref{sec:result}, we find that there are some factors that influence the chemical evolution of galaxies and are reflected in the correlations. The metal fluctuation diagram seems to have a similar structure to the spiral arm in \autoref{fig:final}, this means that the time span between star formation and metal injection is not enough to forget the correlation of the star formation position. This feature that is not captured by the KT18 model may have contributed to the correlation being less than 0 over 6-12 kpc in \autoref{fig:N14_450}. Moreover in \autoref{fig:final}, we find that the fluctuation diagram for $^{12}$C is coarser than for other isotopes, and $^{12}$C appears to be distinctive in various correlation statistics (see auto-correlation in \autoref{fig:corr} and \autoref{fig:corr_value}; temporal correlation in \autoref{fig:time_corr}; cross-correlation in \autoref{fig:cross}). In \autoref{ssec:cross_corr}, we have made a preliminary conclusion that the source of the metal affects the correlation. It forms a picture in which stars are produced centrally in specific regions (spiral arms) and stars of different masses inject various metals into the nearby ISM after different delays (stellar lifetimes). Overall, we are concerned with these two questions: 1. What is the effect of metal injection's stucture on the correlation? 2. What is the effect of the injection delay of different metals on the correlation?
%We will discuss these two most important factors, spiral arm and yield delay, in this section.

In \autoref{sec:result} we saw two major themes in the results. First, there are systematic differences between isotopes in their spatial statistics, temporal statistics, and correlations with each other. In these categories the isotopes we have included appear to separate into three groups. Second, while a simple diffusion model like \citetalias{KT18} appears to describe the zeroth-order spatial statistics, to first order they show significant structure that appears to correlate with the large-scale structures in the galaxy, and they deviate significantly from the \citetalias{KT18} prediction on small scales. We now seek to understand the physical origins of these results.

\subsection{On the statistics of element families}\label{ssec:yield_delay}

We have seen that the spatial statistics of the isotopes we include in our simulation break into three rough groups: $^{12}$C is alone in one group and is the most different from the others, and then the remaining two consist of $^{16}$O, $^{24}$Mg, and $^{32}$S in one group and $^{14}$N, $^{138}$Ba, and $^{140}$Ce in the other. In order to understand the origins of this grouping, it is helpful to examine which stars are responsible for producing which isotopes at which times.

\begin{figure*}
    \includegraphics[width=\textwidth]{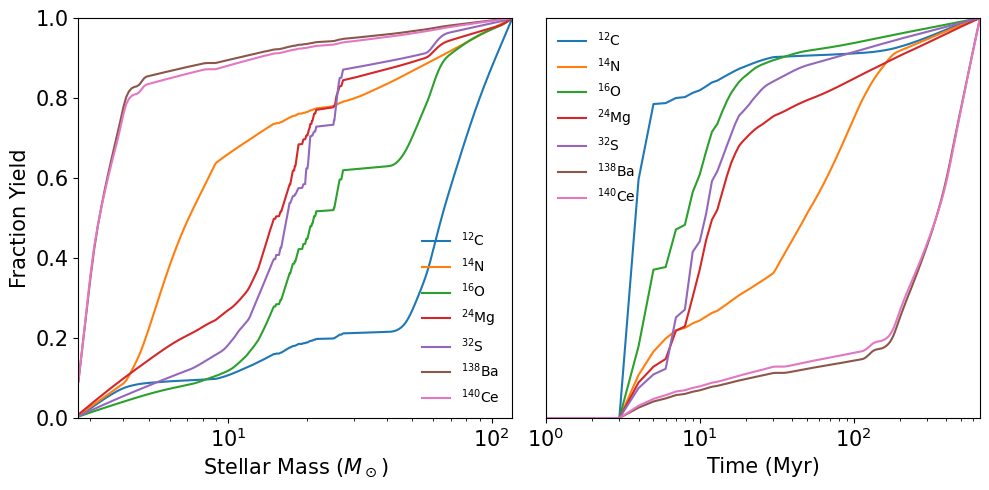}
    \caption{Cumulative yield $M_X(<M,t)$ as a function of initial mass at $t=664$ Myr (\autoref{eq:yield_cum}; left panel) and IMF-integrated yield normalised to the yield at $t=664$ Myr, $M_X(t)/M_X(664\mbox{ Myr})$ (\autoref{eq:yield_tot}; right panel), computed using \textsc{slug} for the same IMF and yield tables used in our simulations. Colours correspond to different isotopes, as indicated in the legend.}
    \label{fig:yield}
\end{figure*}

For an IMF $dN/dM$ and a set of stellar evolution models that predict the total cumulative mass $M_X(M,t)$ of some isotope $X$ that is returned to the ISM by a star of initial mass $M$ and age $t$, we can write the IMF-integrated mass return as a function of time as
\begin{equation}
    M_X(t) = \frac{\int_{M_0}^{M_1} M_X(M,t) \frac{dN}{dM} \, dM}{\int_0^\infty \frac{dN}{dM} \, dM},
    \label{eq:yield_tot}
\end{equation}
where $M_0 = 0.08$ M$_\odot$ to $M_1 = 120$ M$_\odot$ are the minimum and maximum stellar masses for our chosen \citet{Chabrier05a} IMF. Similarly, we can write the cumulative contribution to metal return at age $t$ from stars of mass $<M$ as
\begin{equation}
    M_X(<M, t) = \frac{\int_{M_0}^M M_X(M,t) \frac{dN}{dM}}{\int_{M_0}^{M_1} \frac{dN}{dM} \, dM}.
    \label{eq:yield_cum}
\end{equation}
To evaluate \autoref{eq:yield_tot} for our case, we run \textsc{slug} in its non-stochastic mode (i.e., numerically integrating all quantities over the IMF rather than randomly drawing individual stars) using the same \citet{Chabrier05a} IMF and same set of yield tables as in our simulations, integrating to a maximum time $t = 664$ Myr, the duration of our simulation. Similarly, we evaluate \autoref{eq:yield_cum} by running \textsc{slug} using an IMF that is a $\delta$ function at some initial $M$, again running to $t=664$ Myr and recording the cumulative yield at this time. We carry out this procedure for every mass from $2.7$ M$_\odot$, the minimum stellar mass for which the return is non-zero at $664$ Myr given our choice of stellar tracks and yield tables, to $M_1 = 120$ M$_\odot$ in steps of $0.1$ M$_\odot$. We then numerically evaluate the integral in \autoref{eq:yield_cum}. We plot the cumulative yield $M_X(<M,t)$ at $t=664$ Myr and the IMF-integrated yield $M_X(t)$ as a function of time in the left and right panels of \autoref{fig:yield}. For convenience we can also divide the total yield into three distinct channels: AGB injection from stars with initial masses $2.7 - 8$ M$_\odot$, SN injection by stars from $8-30$ M$_\odot$ and Wolf–Rayet (WR) injection from stars $> 30$ M$_\odot$, though we caution that these divisions are not entirely precise, neither the boundary between stars that do and do not pass through a WR phase nor the boundaries between stars that do and do not explode successfully as SNe lie at a single mass for our chosen tracks and SN models. With this caveat in mind, \autoref{tab:yields} gives the contribution ratio of three nucleosynthesic channels to each of the isotopes we follow and for a stellar population age $t=664$ Myr -- the latter is important, because it means that we are not including a large contribution to carbon from lower-mass stars with lifetimes longer than this; we return to this point below.

\begin{table}\label{tab:AGB-SN-WR}
	\centering
	\caption{Fraction of each each isotope that we follow in our simulations produced via the AGB, SN, and WR nucleosynthesic channels for a stellar population of age $t=664$ Myr aged cluster. Bold numbers highlight the  dominant yield channel for this isotope.
    \label{tab:yields}}
	\begin{tabular}{lccc} % four columns, alignment for each
		\hline
		isotopes & AGB & SN & WR\\
		\hline
		$^{12}$C & 0.10 & 0.11 & \textbf{0.79}\\
		$^{14}$N & \textbf{0.57} & 0.23 & 0.20\\
		$^{16}$O & 0.09 & \textbf{0.53} & 0.38\\
		$^{24}$Mg & 0.23 & \textbf{0.62} & 0.15\\
		$^{32}$S & 0.14 & \textbf{0.74} & 0.12\\
		$^{138}$Ba & \textbf{0.89} & 0.06 & 0.05\\
		$^{140}$Ce & \textbf{0.87} & 0.07 & 0.06\\
		\hline
	\end{tabular}
\end{table}

The most important point to take from \autoref{fig:yield} and \autoref{tab:yields} is that the three natural groupings we found in the spatial statistics of the different isotopes we follow are also visible in the yields as a function of mass and time, and the dominant nucleosynthetic channel. At the age $t=664$ Myr corresponding to the run time of our simulation, $^{12}$C is mostly produced by stars with initial masses $>40$ M$_\odot$ during their WR phase at $t\approx 3-4$ Myr; $^{14}$N, $^{138}$Ba, and $^{140}$Ce are dominated by AGB stars and emerges at stellar ages $\gtrsim 50$ Myr (and even long for Ba and Ce); and $^{16}$O, $^{24}$Mg, and $^{32}$S are mainly contributed by SNe at stellar ages $\approx 10-20$ Myr. 

The clear link between nucleosynthetic site (or equivalently age of stars that produce a particular isotope) and spatial statistics suggests a simple interpretation: in our simulations $^{12}$C is injected mostly by very rare events associated with the formation of the most massive stars, and injection happens almost immediately after these stars form. Because few clusters inject it and because the injection precedes the dispersal of much of the circumstellar gas by SNe, $^{12}$C winds up with both the smallest correlation length and the least correlation with other isotopes. Injection of $^{12}$C is followed by injection of $^{16}$O, $^{24}$Mg, and $^{32}$S, which are all coincident with SNe, which dramatically re-arrange the gas compared to the configuration that existed at the time of $^{12}$C injection. This explains the lower cross-correlation between these elements and $^{12}$C, as well as these elements longer correlation lengths. Finally, $^{14}$N, $^{138}$Ba, and $^{140}$Ce emerge from AGB stars and are not associated with energetic events that re-arrange the gas. These wind up with very slightly larger correlation lengths and somewhat reduced cross-correlations with the SN-associated events because the stars that inject these elements have finite time to drift from their birth sites before giving up their metals, but because their injection is not associated with a dramatic rearrangement of the gas by energetic feedback, the differences are smaller than those between $^{12}$C and the others.

We note that this argument raises the question of whether $^{12}$C would be as different from other elements as we find if we were to continue the simulation for multiple Gyr, such that we could capture the return of C by $\approx 1 - 2.5$ M$_\odot$ stars on these very long timescales. This would presumably shift the behaviour of $^{12}$C somewhat closer to that of the other AGB elements. On the other hand, over such long times the stars responsible for element injection may have so thoroughly phase-mixed that their contribution is effectively azimuthally uniform, in which case this contribution would be important for the total yield but unimportant for spatial fluctuation statistics. Determining where between these two possibilities reality lies will require longer timescale simulations, but maintaining something approaching the high resolution we have available here rather, rather than moving to the much lower resolution used in cosmological simulations.

\subsection{Galactic spiral structure imprinted on metallicity distributions}\label{ssec:spiral_arm}

As illustrated in \autoref{fig:final}, the spiral structure that is visible in the galactic gas and stellar distribution is reflected in both the metallicity and the metallicity fluctuation maps. This structure, is, in turn, likely responsible for the both the non-monotonic behaviour and the negative values that we see in the autocorrelation at some lags (\autoref{fig:N14_450}). The physical origin of this correlation is likely just that spiral arms are where the majority of the stars form, and that supernovae and metal injection occur shortly thereafter. As a result there is a large-scale pattern imprinted on metal injection, which in turn leads to a similar large-scale pattern on the fluctuation distribution. 

\begin{figure}
    \includegraphics[width=\columnwidth]{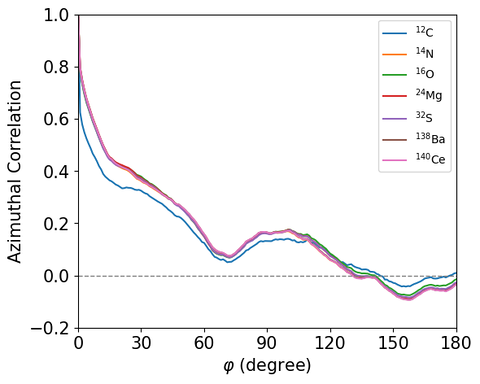}
    \caption{Azimuthal correlations $\xi_\mathrm{az}(\varphi)$ (\autoref{eq:corr_az}) as a function of rotation angle $\varphi$ at 564 Myr. Colours correspond to different isotopes, as indicated in the legend.}
    \label{fig:ro_564}
\end{figure}

To demonstrate that these effects are indeed a result of spiral structure, we can instead of studying the full autocorrelation, which involves translations of the map in arbitrary directions, instead study the autocorrelation under rotations above the galactic centre. To be precise, we define the azimuthal autocorrelation of a metallicity fluctuation field $Z'$ as
\begin{equation}
    \xi_\mathrm{az}(\varphi) =  \frac{\left\langle Z'(\mathbf{r}) Z'(\mathbf{r} + \varphi \, \hat{\varphi})\right\rangle_\mathbf{r}}{\left\langle Z'^2(\mathbf{r})\right\rangle_\mathbf{r}},
    \label{eq:corr_az}
\end{equation}
where $\langle\cdot\rangle_\mathbf{r}$ indicates an average over our usual $15\mbox{ kpc}\times 15\mbox{ kpc}$ region centred on the galactic centre at $\mathbf{r} = 0$. Intuitively, $\xi_\mathrm{az}$ simply measures the strength of the correlation between the metal field and a version of the metal field that has been rotated through an angle $\varphi$. This quantity is interesting because of its ability to pick out the presence of large-scale rotationally-symmetric structures such as spiral arms. In the absence of such structures, we would expect \(\xi_\mathrm{az}(\varphi)\) to approach a constant, non-negative value for large rotation angles. However, if the metal field contains large-scale rotationally symmetric structures, there will be some rotation angles where regions of high $Z'$ in the rotated and non-rotated maps overlap, leading to a high correlation, and other rotation angles where high $Z'$ regions in the rotated map preferentially align with low $Z'$ regions in the non-rotated map, yielding a negative correlation. We plot $\xi_\mathrm{az}(\varphi)$ at $t=564$ Myr in \autoref{fig:ro_564}. We see a positive but monotonically decreasing correlation from 0 to 70 degrees, a small bump at roughly 70 to 100 degrees, and then a negative correlation at 100 to 180 degrees. The presence of these features is consistent with the idea that the non-monotonic behaviour we observe in the full spatial autocorrelation (\autoref{fig:corr}) is a signature of spiral features with large-scale rotational symmetry.

%Similar spiral structures in the metallicity fluctuation maps from simulated galaxies \citep{2021MNRAS.505.4586B,2023MNRAS.521.3708O}, and there is observational evidence for such structures in nearby galaxies \citep{Ho18a} and in the Milky Way \citep{2019ApJ...887..114W}. 

Spiral-induced features in gas-phase metallicity distributions have not been explored quantitatively before in theoretical models and simulations. The \citetalias{KT18} model explicitly assumes that metal injection is a Poissonian process, with no spatial structure. Previous N-body simulations of isolated galaxies have found spiral arm-associated variations in \textit{stellar} metallicity that they have attributed to stellar radial motions induced by spiral arms \citep{Grand15b, 2016MNRAS.460L..94G, 2016ApJ...830L..40S}, but this mechanism depends on the ability of collisionless stars to move radially inward and outward in response to gravitational perturbations from a passing arm. It is clearly inapplicable to the gas. Cosmological simulations have also found evidence for spiral features in gas-phase metallicity maps \citep{2021MNRAS.505.4586B,2023MNRAS.521.3708O,2024MNRAS.528.7103L}, but at much lower resolution than in our work, and the authors of these works either have not searched for spiral effects on metallicity correlation functions, or have not found convincing evidence for them (e.g., compare Figure 3 of \citealt{2024MNRAS.528.7103L} to our \autoref{fig:N14_450}). Thus our simulations appear to the first to identify spiral structures in higher-order statistics.

The fact that we see these patterns while previous simulations have not may be a resolution effect: the previous simulations that have found spiral features, from the FIRE-2 and Auriga projects, have mass resolutions of $\gtrsim 5\times 10^3$ M$_\odot$, more than an order of magnitude lower than our $\approx 300$ M$_\odot$. At the mean density of the Milky Way's ISM, $\approx 1$ cm$^{-3}$, the former corresponds to $\approx 60$ pc resolution, while the latter corresponds to $\approx 20$ pc. This difference means that the spiral arms in our simulations are significantly sharper than in the Auriga or FIRE-2 simulations, which may be why we see a stronger spiral arm effect.

Interestingly, there does appear to be some observational evidence for spiral-aligned features in high-resolution metallicity maps. \citet{2016ApJ...830L..40S} find evidence for spiral structure in high-resolution metallicity fluctuation maps derived from MUSE observations of NGC 6754. \citet{Ho17a, 2018A&A...618A..64H} report similar structures in two galaxies observed as part of the TYPHOON survey, and \citet{2019ApJ...887..114W} find evidence for spiral features in the metallicity distribution of H~\textsc{ii} regions in the Milky Way. The features we identify here appear to be at least qualitatively consistent with these observations.

\subsection{The injection width and small-scale structures}
\label{ssec:winj}

In addition to the large-scale features in the elemental auto-correlation imprinted by galactic-scale structures like spiral arms, we also see behaviour at small scales that is not fully captured by models. In the context of the original \citetalias{KT18} model, the injection width is the effective radius over which a single event injects metals, before those metals begin to be transported and mixed by general ISM turbulence. If this view were correct, we would expect much smaller injection widths for the isotopes that come predominantly from AGB stars, where injection is not accompanied by an explosive release of energy, than for those injected by primarily by SNe, where it is. By contrast, the correlation length is assumed to be reflective of the properties of ISM turbulence, and thus should presumably be similar for all elements. Neither of these assumptions are fully borne out by our data: $^{12}$C, injected by WR stars, has both the smallest correlation length and injection width, while the AGB-dominated isotopes that we follow ($^{14}$N, $^{138}$Ba, and $^{140}$Ce) generally have slightly larger injection widths that the SN-dominated ones ($^{16}$O, $^{24}$Mg, $^{32}$S).

It is suggestive that, while these isotopes' injection widths do not appear to be correlated with the present of absence of energetic events at the time of injection, they do appear to form a sequence in time: the injection width is smallest for the isotope that is injected first (and prior to the onset of SNe) at stellar population ages of $\approx 3-4$ Myr -- $^{12}$C -- and largest for the AGB-produced isotopes that are injected on the longest timescales of $\sim 100$ Myr. The SN-injected ones injected at $\sim 10-20$ Myr sit between these two. One possible explanation for this finding is that the primary determinant of injection width is not so much the size of the bubble blown by the energy accompany element return, but whether injection happens prior to dispersal of star-forming clouds, and -- closely related -- whether stars and gas have time to move relative to one another prior to element returns. The return of significant $^{12}$C on short timescales makes it unique in that return happens before either SNe or stellar drift have had time to induce significant gas-star separation -- although our simulations include photoionisation, \citet{Jeffreson24a} point out that this is insufficient to disperse massive molecular clouds, and these these massive clouds disproportionately drive galactic-scale correlations of SNe. It seems reasonable to hypothesise that the same is true for element distributions. By contrast, SNe-injected elements necessarily accompany gas dispersal, and AGB elements are injected after both gas dispersal and significant drift of stars away from their birth sites have taken place, and thus are injected over an even wider area than SN-produced ones. Such a picture is consistent with the ordering of injection widths that we have uncovered in our simulations.

\section{Conclusion}\label{sec:conclusion}

We carry out high-resolution simulations of an isolated, Milky Way-like galaxy in order to provide comprehensive insights into the mechanisms that govern the spatial and temporal distribution of metals in galaxies. By simulating the injection of key isotopes -- such as $^{12}$C, $^{14}$N, $^{16}$O, $^{24}$Mg, $^{32}$S, $^{138}$Ba, and $^{140}$Ce -- using a detailed star-by-star feedback approach, we have been able to observe how different nucleosynthetic processes contribute to galactic metal distributions, and how these distributions reflect elements' diverse nucleosynthetic origins. Our simulations are the first to study this question at a resolution sufficient to capture the vertical structure and thus the turbulence in the interstellar medium.

We find that the metallicity fluctuation distributions -- the residuals left once we remove the overall radial metallicity gradient -- of all the isotopes settle to statistical steady state over timescales of $\sim 200$ Myr, roughly one galactic rotation. In this steady state the metallicty distributions are correlated on $\sim \mathrm{kpc}$ scales, comparable to the correlation lengths observed in nearby Milky Way-mass galaxies \citep[e.g.,][]{2021MNRAS.504.5496L, 2023MNRAS.518..286L}. In this steady state the spatial auto-correlation functions of element spatial distributions are reasonably well-described to zeroth order by the predictions of simple injection-diffusion models such as that proposed by \citet{KT18}, and the spatial patterns of the metallicity field remain correlated on timescales of $\sim 100$ Myr. The abundances of different isotopes are also highly correlated with one another, such that the effective dimensonality of the chemical element space spanned by young stars is likely to be much smaller than one might have guessed under a naive assumption that correlations in different elements -- or even different element groups -- are independent.

On top of this zeroth-order picture, however, our very high resolution allows us to detect significant additional structure in the spatial statistics of isotope distributions. The different isotopes we follow naturally fall into three groups depending on their dominant nucleosynthetic origins.
Specifically, $^{12}$C, which over the $\approx 0.7$ Gyr duration of our simulations is primarily produced by Wolf-Rayet stars, has the smallest spatial correlation length, shortest temporal correlation, and is most weakly correlated with the other elements we follow. These features are likely a result of its very rapid injection, with both precedes the onset of supernovae and occurs before stars have time to undergo significant drift relative to the gas from which they formed. The isotopes we follow that are primarily produced by AGB stars -- $^{14}$N, $^{138}$Ba, and $^{140}$Ce -- have the longest correlation lengths and are extremely well correlated with one another, likely because the stars that produce them have plenty of time to disperse before they die. Finally, the isotopes we follow whose nucleosynthetic origin is primarily in type II supernovae -- $^{16}$O, $^{24}$Mg, and $^{32}$S -- have an intermediate correlation length and correlate best with each other, slightly less with AGB-produced elements, and still less with $^{12}$C, although we emphasise that all of these correlations are still strong in an absolute sense. The different nucleosynthetic origin sites also affect the small-scale structure of the spatial correlation, the quantity that is described in the \citet{KT18} model as the ``injection width'' that characterises the size of the region into which each stellar source deposits its metals before they begin to be mixed into the ISM. In the 
\citeauthor{KT18} this size scale is expected to be determined by the amount of energy that accompanies element release, so supernovae, which inflate large bubbles in the ISM, should have large injection widths than AGB stars, which do not. We find that this expectation is not satisfied in our simulations, and that AGB-produced elements are more correlated on small scales (i.e., have larger injection width) than other isotopes. This suggests that stellar drift and the timing of element injection relative to supernovae is a dominant factor in establishing the small-scale behaviour of metallicity correlations. 

In addition to these isotope-by-isotope analyses, thanks to our high resolution that captures spiral features in galaxies very well, we are able for the first time to see clear imprints of these galactic-scale structure in statistics of galactic gas-phase metallicity distributions. Spiral structures lead to large-scale non-monotonic and oscillatory patterns in the autocorrelations of metallicity distributions that differ from those predicted by simpler diffusion models. The origin of these features is characteristic fluctuations of increased metallicity that align with regions of enhanced star formation, and the existence of such features underscores the need to incorporate the effects of galactic dynamics when interpreting observed metallicity maps.

Future research should continue to explore the impact of different feedback processes, improve the modeling of stellar drift, and integrate these findings with observational data to enhance our understanding of chemical evolution in galaxies. In addition to increasing the number of distinct isotopes that we follow overall, incorporating feedback and metal injection from Type Ia supernovae and neutron stars mergers will also be crucial to extending our statistical analysis to include the iron peak and r-process elements that are primarily produced by events of these types. 
%Additionally, we could set up isolated metal fields in the simulations that mix alongside the global metal field. This approach could more effectively analyze the contribution of each nucleosynthetic process to the overall chemical evolution. 
Another substantial upgrade that we intend to include in future work is for newborn stars to inherit the chemical information of their gas particles, allowing us to study the spatial and element-to-element correlations of the stars themselves, and the coupling between stars and gas, following the stars past the moment of their formation. Doing so will be crucial to understanding how stars of that are born near to each other and thus have similar chemical compositions disperse in space, broadening the distribution of abundance patterns at fixed position. Understanding how this process happens is a crucial step toward efforts at chemical tagging -- using stellar abundances to reconstruct the star formation history of galaxies \citep[e.g.,][]{Bland-Hawthorn16a,Krumholz19a}. By incorporating both observational insights and theoretical models, we can refine our understanding of how metals evolve in galaxies, ultimately providing a more complete picture of galactic evolution and star formation processes.

\section*{Acknowledgements}

The authors acknowledge helpful conversations with Yuan-Sen Ting, Melissa Ness, and Trey Wenger. We acknowledge high-performance computing resources provided by the Australian National Computational Infrastructure (award jh2) through the National and ANU Computational Merit Allocation Schemes. ZL acknowledges the Science and Technology Facilities Council (STFC) consolidated grant
ST/X001075/1. MRK acknowledges support from the Australian Research Council through Laureate Fellowship FL220100020.

%%%%%%%%%%%%%%%%%%%%%%%%%%%%%%%%%%%%%%%%%%%%%%%%%%
\section*{Data Availability}

The data underlying this article will be shared upon reasonable request to the corresponding author.



\bibliographystyle{mnras}
\bibliography{example} % if your bibtex file is called example.bib

\begin{thebibliography}{}
\makeatletter
\relax
\def\mn@urlcharsother{\let\do\@makeother \do\$\do\&\do\#\do\^\do\_\do\%\do\~}
\def\mn@doi{\begingroup\mn@urlcharsother \@ifnextchar [ {\mn@doi@} {\mn@doi@[]}}
\def\mn@doi@[#1]#2{\def\@tempa{#1}\ifx\@tempa\@empty \href {http://dx.doi.org/#2} {doi:#2}\else \href {http://dx.doi.org/#2} {#1}\fi \endgroup}
\def\mn@eprint#1#2{\mn@eprint@#1:#2::\@nil}
\def\mn@eprint@arXiv#1{\href {http://arxiv.org/abs/#1} {{\tt arXiv:#1}}}
\def\mn@eprint@dblp#1{\href {http://dblp.uni-trier.de/rec/bibtex/#1.xml} {dblp:#1}}
\def\mn@eprint@#1:#2:#3:#4\@nil{\def\@tempa {#1}\def\@tempb {#2}\def\@tempc {#3}\ifx \@tempc \@empty \let \@tempc \@tempb \let \@tempb \@tempa \fi \ifx \@tempb \@empty \def\@tempb {arXiv}\fi \@ifundefined {mn@eprint@\@tempb}{\@tempb:\@tempc}{\expandafter \expandafter \csname mn@eprint@\@tempb\endcsname \expandafter{\@tempc}}}

\bibitem[\protect\citeauthoryear{{Armillotta}, {Krumholz}, {Di Teodoro}  \& {McClure-Griffiths}}{{Armillotta} et~al.}{2019}]{2019MNRAS.490.4401A}
{Armillotta} L.,  {Krumholz} M.~R.,  {Di Teodoro} E.~M.,   {McClure-Griffiths} N.~M.,  2019, \mn@doi [\mnras] {10.1093/mnras/stz2880}, \href {https://ui.adsabs.harvard.edu/abs/2019MNRAS.490.4401A} {490, 4401}

\bibitem[\protect\citeauthoryear{{Belfiore} et~al.,}{{Belfiore} et~al.}{2017}]{2017MNRAS.469..151B}
{Belfiore} F.,  et~al., 2017, \mn@doi [\mnras] {10.1093/mnras/stx789}, \href {https://ui.adsabs.harvard.edu/abs/2017MNRAS.469..151B} {469, 151}

\bibitem[\protect\citeauthoryear{{Bellardini}, {Wetzel}, {Loebman}, {Faucher-Gigu{\`e}re}, {Ma}  \& {Feldmann}}{{Bellardini} et~al.}{2021}]{2021MNRAS.505.4586B}
{Bellardini} M.~A.,  {Wetzel} A.,  {Loebman} S.~R.,  {Faucher-Gigu{\`e}re} C.-A.,  {Ma} X.,   {Feldmann} R.,  2021, \mn@doi [\mnras] {10.1093/mnras/stab1606}, \href {https://ui.adsabs.harvard.edu/abs/2021MNRAS.505.4586B} {505, 4586}

\bibitem[\protect\citeauthoryear{{Bland-Hawthorn} \& {Sharma}}{{Bland-Hawthorn} \& {Sharma}}{2016}]{Bland-Hawthorn16a}
{Bland-Hawthorn} J.,  {Sharma} S.,  2016, \mn@doi [Astronomische Nachrichten] {10.1002/asna.201612393}, \href {http://adsabs.harvard.edu/abs/2016AN....337..894B} {337, 894}

\bibitem[\protect\citeauthoryear{{Bland-Hawthorn}, {Krumholz}  \& {Freeman}}{{Bland-Hawthorn} et~al.}{2010}]{Bland-Hawthorn10a}
{Bland-Hawthorn} J.,  {Krumholz} M.~R.,   {Freeman} K.,  2010, \mn@doi [\apj] {10.1088/0004-637X/713/1/166}, \href {http://adsabs.harvard.edu/abs/2010ApJ...713..166B} {713, 166}

\bibitem[\protect\citeauthoryear{{Bressan}, {Marigo}, {Girardi}, {Salasnich}, {Dal Cero}, {Rubele}  \& {Nanni}}{{Bressan} et~al.}{2012}]{2012MNRAS.427..127B}
{Bressan} A.,  {Marigo} P.,  {Girardi} L.,  {Salasnich} B.,  {Dal Cero} C.,  {Rubele} S.,   {Nanni} A.,  2012, \mn@doi [\mnras] {10.1111/j.1365-2966.2012.21948.x}, \href {https://ui.adsabs.harvard.edu/abs/2012MNRAS.427..127B} {427, 127}

\bibitem[\protect\citeauthoryear{{Bundy} et~al.,}{{Bundy} et~al.}{2015}]{2015ApJ...798....7B}
{Bundy} K.,  et~al., 2015, \mn@doi [\apj] {10.1088/0004-637X/798/1/7}, \href {https://ui.adsabs.harvard.edu/abs/2015ApJ...798....7B} {798, 7}

\bibitem[\protect\citeauthoryear{{Ceverino}, {S{\'a}nchez Almeida}, {Mu{\~n}oz Tu{\~n}{\'o}n}, {Dekel}, {Elmegreen}, {Elmegreen}  \& {Primack}}{{Ceverino} et~al.}{2016}]{2016MNRAS.457.2605C}
{Ceverino} D.,  {S{\'a}nchez Almeida} J.,  {Mu{\~n}oz Tu{\~n}{\'o}n} C.,  {Dekel} A.,  {Elmegreen} B.~G.,  {Elmegreen} D.~M.,   {Primack} J.,  2016, \mn@doi [\mnras] {10.1093/mnras/stw064}, \href {https://ui.adsabs.harvard.edu/abs/2016MNRAS.457.2605C} {457, 2605}

\bibitem[\protect\citeauthoryear{{Chabrier}}{{Chabrier}}{2005}]{Chabrier05a}
{Chabrier} G.,  2005, in {Corbelli} E.,  {Palla} F.,   {Zinnecker} H.,  eds,  Astrophysics and Space Science Library Vol. 327, The Initial Mass Function 50 Years Later. p.~41 (\mn@eprint {arXiv} {astro-ph/0409465}), \mn@doi{10.1007/978-1-4020-3407-7_5}

\bibitem[\protect\citeauthoryear{{Colbrook}, {Ma}, {Hopkins}  \& {Squire}}{{Colbrook} et~al.}{2017}]{2017MNRAS.467.2421C}
{Colbrook} M.~J.,  {Ma} X.,  {Hopkins} P.~F.,   {Squire} J.,  2017, \mn@doi [\mnras] {10.1093/mnras/stx261}, \href {https://ui.adsabs.harvard.edu/abs/2017MNRAS.467.2421C} {467, 2421}

\bibitem[\protect\citeauthoryear{{Croom} et~al.,}{{Croom} et~al.}{2012}]{2012MNRAS.421..872C}
{Croom} S.~M.,  et~al., 2012, \mn@doi [\mnras] {10.1111/j.1365-2966.2011.20365.x}, \href {https://ui.adsabs.harvard.edu/abs/2012MNRAS.421..872C} {421, 872}

\bibitem[\protect\citeauthoryear{{Di Matteo}, {Pipino}, {Lehnert}, {Combes}  \& {Semelin}}{{Di Matteo} et~al.}{2009}]{2009A&A...499..427D}
{Di Matteo} P.,  {Pipino} A.,  {Lehnert} M.~D.,  {Combes} F.,   {Semelin} B.,  2009, \mn@doi [\aap] {10.1051/0004-6361/200911715}, \href {https://ui.adsabs.harvard.edu/abs/2009A&A...499..427D} {499, 427}

\bibitem[\protect\citeauthoryear{{Di Matteo}, {Haywood}, {Combes}, {Semelin}  \& {Snaith}}{{Di Matteo} et~al.}{2013}]{2013A&A...553A.102D}
{Di Matteo} P.,  {Haywood} M.,  {Combes} F.,  {Semelin} B.,   {Snaith} O.~N.,  2013, \mn@doi [\aap] {10.1051/0004-6361/201220539}, \href {https://ui.adsabs.harvard.edu/abs/2013A&A...553A.102D} {553, A102}

\bibitem[\protect\citeauthoryear{{Doherty}, {Gil-Pons}, {Lau}, {Lattanzio}  \& {Siess}}{{Doherty} et~al.}{2014}]{2014MNRAS.437..195D}
{Doherty} C.~L.,  {Gil-Pons} P.,  {Lau} H. H.~B.,  {Lattanzio} J.~C.,   {Siess} L.,  2014, \mn@doi [\mnras] {10.1093/mnras/stt1877}, \href {https://ui.adsabs.harvard.edu/abs/2014MNRAS.437..195D} {437, 195}

\bibitem[\protect\citeauthoryear{{Emsellem} et~al.,}{{Emsellem} et~al.}{2022}]{2022A&A...659A.191E}
{Emsellem} E.,  et~al., 2022, \mn@doi [\aap] {10.1051/0004-6361/202141727}, \href {https://ui.adsabs.harvard.edu/abs/2022A&A...659A.191E} {659, A191}

\bibitem[\protect\citeauthoryear{{Erroz-Ferrer} et~al.,}{{Erroz-Ferrer} et~al.}{2019}]{2019MNRAS.484.5009E}
{Erroz-Ferrer} S.,  et~al., 2019, \mn@doi [\mnras] {10.1093/mnras/stz194}, \href {https://ui.adsabs.harvard.edu/abs/2019MNRAS.484.5009E} {484, 5009}

\bibitem[\protect\citeauthoryear{{Foreman-Mackey}, {Hogg}, {Lang}  \& {Goodman}}{{Foreman-Mackey} et~al.}{2013}]{2013PASP..125..306F}
{Foreman-Mackey} D.,  {Hogg} D.~W.,  {Lang} D.,   {Goodman} J.,  2013, \mn@doi [\pasp] {10.1086/670067}, \href {https://ui.adsabs.harvard.edu/abs/2013PASP..125..306F} {125, 306}

\bibitem[\protect\citeauthoryear{{Grand}, {Bovy}, {Kawata}, {Hunt}, {Famaey}, {Siebert}, {Monari}  \& {Cropper}}{{Grand} et~al.}{2015}]{Grand15b}
{Grand} R. J.~J.,  {Bovy} J.,  {Kawata} D.,  {Hunt} J. A.~S.,  {Famaey} B.,  {Siebert} A.,  {Monari} G.,   {Cropper} M.,  2015, \mn@doi [\mnras] {10.1093/mnras/stv1785}, \href {https://ui.adsabs.harvard.edu/abs/2015MNRAS.453.1867G} {453, 1867}

\bibitem[\protect\citeauthoryear{{Grand} et~al.,}{{Grand} et~al.}{2016}]{2016MNRAS.460L..94G}
{Grand} R. J.~J.,  et~al., 2016, \mn@doi [\mnras] {10.1093/mnrasl/slw086}, \href {https://ui.adsabs.harvard.edu/abs/2016MNRAS.460L..94G} {460, L94}

\bibitem[\protect\citeauthoryear{{Grand} et~al.,}{{Grand} et~al.}{2017}]{2017MNRAS.467..179G}
{Grand} R. J.~J.,  et~al., 2017, \mn@doi [\mnras] {10.1093/mnras/stx071}, \href {https://ui.adsabs.harvard.edu/abs/2017MNRAS.467..179G} {467, 179}

\bibitem[\protect\citeauthoryear{Grudić}{Grudić}{2021}]{Grudic_meshoid_2021}
Grudić M.~Y.,  2021, {meshoid}, \url {https://github.com/mikegrudic/meshoid}

\bibitem[\protect\citeauthoryear{{Ho} et~al.,}{{Ho} et~al.}{2017}]{Ho17a}
{Ho} I.~T.,  et~al., 2017, \mn@doi [\apj] {10.3847/1538-4357/aa8460}, \href {https://ui.adsabs.harvard.edu/abs/2017ApJ...846...39H} {846, 39}

\bibitem[\protect\citeauthoryear{{Ho} et~al.,}{{Ho} et~al.}{2018}]{2018A&A...618A..64H}
{Ho} I.~T.,  et~al., 2018, \mn@doi [\aap] {10.1051/0004-6361/201833262}, \href {https://ui.adsabs.harvard.edu/abs/2018A&A...618A..64H} {618, A64}

\bibitem[\protect\citeauthoryear{{Hopkins}}{{Hopkins}}{2015}]{2015MNRAS.450...53H}
{Hopkins} P.~F.,  2015, \mn@doi [\mnras] {10.1093/mnras/stv195}, \href {https://ui.adsabs.harvard.edu/abs/2015MNRAS.450...53H} {450, 53}

\bibitem[\protect\citeauthoryear{{Hopkins} et~al.,}{{Hopkins} et~al.}{2018}]{2018MNRAS.480..800H}
{Hopkins} P.~F.,  et~al., 2018, \mn@doi [\mnras] {10.1093/mnras/sty1690}, \href {https://ui.adsabs.harvard.edu/abs/2018MNRAS.480..800H} {480, 800}

\bibitem[\protect\citeauthoryear{{Hu}, {Wibking}  \& {Krumholz}}{{Hu} et~al.}{2023}]{Hu23}
{Hu} Z.,  {Wibking} B.~D.,   {Krumholz} M.~R.,  2023, \mn@doi [\mnras] {10.1093/mnras/stad931}, \href {https://ui.adsabs.harvard.edu/abs/2023MNRAS.521.5604H} {521, 5604}

\bibitem[\protect\citeauthoryear{{Jeffreson}, {Semenov}  \& {Krumholz}}{{Jeffreson} et~al.}{2024}]{Jeffreson24a}
{Jeffreson} S. M.~R.,  {Semenov} V.~A.,   {Krumholz} M.~R.,  2024, \mn@doi [\mnras] {10.1093/mnras/stad3550}, \href {https://ui.adsabs.harvard.edu/abs/2024MNRAS.527.7093J} {527, 7093}

\bibitem[\protect\citeauthoryear{{Karakas} \& {Lugaro}}{{Karakas} \& {Lugaro}}{2016}]{2016ApJ...825...26K}
{Karakas} A.~I.,  {Lugaro} M.,  2016, \mn@doi [\apj] {10.3847/0004-637X/825/1/26}, \href {https://ui.adsabs.harvard.edu/abs/2016ApJ...825...26K} {825, 26}

\bibitem[\protect\citeauthoryear{{Kewley}, {Nicholls}  \& {Sutherland}}{{Kewley} et~al.}{2019}]{Kewley_2019}
{Kewley} L.~J.,  {Nicholls} D.~C.,   {Sutherland} R.~S.,  2019, \mn@doi [\araa] {10.1146/annurev-astro-081817-051832}, \href {https://ui.adsabs.harvard.edu/abs/2019ARA&A..57..511K} {57, 511}

\bibitem[\protect\citeauthoryear{{Kreckel} et~al.,}{{Kreckel} et~al.}{2019}]{2019ApJ...887...80K}
{Kreckel} K.,  et~al., 2019, \mn@doi [\apj] {10.3847/1538-4357/ab5115}, \href {https://ui.adsabs.harvard.edu/abs/2019ApJ...887...80K} {887, 80}

\bibitem[\protect\citeauthoryear{{Kreckel} et~al.,}{{Kreckel} et~al.}{2020}]{2020MNRAS.499..193K}
{Kreckel} K.,  et~al., 2020, \mn@doi [\mnras] {10.1093/mnras/staa2743}, \href {https://ui.adsabs.harvard.edu/abs/2020MNRAS.499..193K} {499, 193}

\bibitem[\protect\citeauthoryear{{Krumholz} \& {Ting}}{{Krumholz} \& {Ting}}{2018}]{KT18}
{Krumholz} M.~R.,  {Ting} Y.-S.,  2018, \mn@doi [\mnras] {10.1093/mnras/stx3286}, \href {https://ui.adsabs.harvard.edu/abs/2018MNRAS.475.2236K} {475, 2236}

\bibitem[\protect\citeauthoryear{{Krumholz}, {Fumagalli}, {da Silva}, {Rendahl}  \& {Parra}}{{Krumholz} et~al.}{2015}]{2015MNRAS.452.1447K}
{Krumholz} M.~R.,  {Fumagalli} M.,  {da Silva} R.~L.,  {Rendahl} T.,   {Parra} J.,  2015, \mn@doi [\mnras] {10.1093/mnras/stv1374}, \href {https://ui.adsabs.harvard.edu/abs/2015MNRAS.452.1447K} {452, 1447}

\bibitem[\protect\citeauthoryear{{Krumholz}, {McKee}  \& {Bland -Hawthorn}}{{Krumholz} et~al.}{2019}]{Krumholz19a}
{Krumholz} M.~R.,  {McKee} C.~F.,   {Bland -Hawthorn} J.,  2019, \mn@doi [\araa] {10.1146/annurev-astro-091918-104430}, \href {https://ui.adsabs.harvard.edu/abs/2019ARA&A..57..227K} {57, 227}

\bibitem[\protect\citeauthoryear{{Leitherer} et~al.,}{{Leitherer} et~al.}{1999}]{1999ApJS..123....3L}
{Leitherer} C.,  et~al., 1999, \mn@doi [\apjs] {10.1086/313233}, \href {https://ui.adsabs.harvard.edu/abs/1999ApJS..123....3L} {123, 3}

\bibitem[\protect\citeauthoryear{{Li}, {Krumholz}, {Wisnioski}, {Mendel}, {Kewley}, {S{\'a}nchez}  \& {Galbany}}{{Li} et~al.}{2021}]{2021MNRAS.504.5496L}
{Li} Z.,  {Krumholz} M.~R.,  {Wisnioski} E.,  {Mendel} J.~T.,  {Kewley} L.~J.,  {S{\'a}nchez} S.~F.,   {Galbany} L.,  2021, \mn@doi [\mnras] {10.1093/mnras/stab1263}, \href {https://ui.adsabs.harvard.edu/abs/2021MNRAS.504.5496L} {504, 5496}

\bibitem[\protect\citeauthoryear{{Li} et~al.,}{{Li} et~al.}{2023}]{2023MNRAS.518..286L}
{Li} Z.,  et~al., 2023, \mn@doi [\mnras] {10.1093/mnras/stac3028}, \href {https://ui.adsabs.harvard.edu/abs/2023MNRAS.518..286L} {518, 286}

\bibitem[\protect\citeauthoryear{{Li}, {Li}, {Wisnioski}, {Krumholz}  \& {S{\'a}nchez}}{{Li} et~al.}{2024a}]{2024arXiv240704252L}
{Li} S.-l.,  {Li} Z.,  {Wisnioski} E.,  {Krumholz} M.~R.,   {S{\'a}nchez} S.~F.,  2024a, \mn@doi [arXiv e-prints] {10.48550/arXiv.2407.04252}, \href {https://ui.adsabs.harvard.edu/abs/2024arXiv240704252L} {p. arXiv:2407.04252}

\bibitem[\protect\citeauthoryear{{Li} et~al.,}{{Li} et~al.}{2024b}]{2024MNRAS.528.7103L}
{Li} Z.,  et~al., 2024b, \mn@doi [\mnras] {10.1093/mnras/stae480}, \href {https://ui.adsabs.harvard.edu/abs/2024MNRAS.528.7103L} {528, 7103}

\bibitem[\protect\citeauthoryear{{L{\'o}pez-Cob{\'a}} et~al.,}{{L{\'o}pez-Cob{\'a}} et~al.}{2020}]{2020AJ....159..167L}
{L{\'o}pez-Cob{\'a}} C.,  et~al., 2020, \mn@doi [\aj] {10.3847/1538-3881/ab7848}, \href {https://ui.adsabs.harvard.edu/abs/2020AJ....159..167L} {159, 167}

\bibitem[\protect\citeauthoryear{{Ma}, {Hopkins}, {Feldmann}, {Torrey}, {Faucher-Gigu{\`e}re}  \& {Kere{\v{s}}}}{{Ma} et~al.}{2017}]{2017MNRAS.466.4780M}
{Ma} X.,  {Hopkins} P.~F.,  {Feldmann} R.,  {Torrey} P.,  {Faucher-Gigu{\`e}re} C.-A.,   {Kere{\v{s}}} D.,  2017, \mn@doi [\mnras] {10.1093/mnras/stx034}, \href {https://ui.adsabs.harvard.edu/abs/2017MNRAS.466.4780M} {466, 4780}

\bibitem[\protect\citeauthoryear{Maiolino \& Mannucci}{Maiolino \& Mannucci}{2019}]{Maiolino_2019}
Maiolino R.,  Mannucci F.,  2019, \mn@doi [The Astronomy and Astrophysics Review] {10.1007/s00159-018-0112-2}, 27

\bibitem[\protect\citeauthoryear{{M{\'a}rmol-Queralt{\'o}} et~al.,}{{M{\'a}rmol-Queralt{\'o}} et~al.}{2011}]{2011A&A...534A...8M}
{M{\'a}rmol-Queralt{\'o}} E.,  et~al., 2011, \mn@doi [\aap] {10.1051/0004-6361/201117032}, \href {https://ui.adsabs.harvard.edu/abs/2011A&A...534A...8M} {534, A8}

\bibitem[\protect\citeauthoryear{{Metha}, {Trenti}  \& {Chu}}{{Metha} et~al.}{2021}]{2021MNRAS.508..489M}
{Metha} B.,  {Trenti} M.,   {Chu} T.,  2021, \mn@doi [\mnras] {10.1093/mnras/stab2554}, \href {https://ui.adsabs.harvard.edu/abs/2021MNRAS.508..489M} {508, 489}

\bibitem[\protect\citeauthoryear{{Orr} et~al.,}{{Orr} et~al.}{2023}]{2023MNRAS.521.3708O}
{Orr} M.~E.,  et~al., 2023, \mn@doi [\mnras] {10.1093/mnras/stad676}, \href {https://ui.adsabs.harvard.edu/abs/2023MNRAS.521.3708O} {521, 3708}

\bibitem[\protect\citeauthoryear{{Petit}, {Krumholz}, {Goldbaum}  \& {Forbes}}{{Petit} et~al.}{2015}]{2015MNRAS.449.2588P}
{Petit} A.~C.,  {Krumholz} M.~R.,  {Goldbaum} N.~J.,   {Forbes} J.~C.,  2015, \mn@doi [\mnras] {10.1093/mnras/stv493}, \href {https://ui.adsabs.harvard.edu/abs/2015MNRAS.449.2588P} {449, 2588}

\bibitem[\protect\citeauthoryear{{Poetrodjojo} et~al.,}{{Poetrodjojo} et~al.}{2018}]{2018MNRAS.479.5235P}
{Poetrodjojo} H.,  et~al., 2018, \mn@doi [\mnras] {10.1093/mnras/sty1782}, \href {https://ui.adsabs.harvard.edu/abs/2018MNRAS.479.5235P} {479, 5235}

\bibitem[\protect\citeauthoryear{{Richard} et~al.,}{{Richard} et~al.}{2024}]{BlueMUSE}
{Richard} J.,  et~al., 2024, \mn@doi [arXiv e-prints] {10.48550/arXiv.2406.13914}, \href {https://ui.adsabs.harvard.edu/abs/2024arXiv240613914R} {p. arXiv:2406.13914}

\bibitem[\protect\citeauthoryear{{Rosales-Ortega}, {D{\'\i}az}, {Kennicutt}  \& {S{\'a}nchez}}{{Rosales-Ortega} et~al.}{2011}]{2011MNRAS.415.2439R}
{Rosales-Ortega} F.~F.,  {D{\'\i}az} A.~I.,  {Kennicutt} R.~C.,   {S{\'a}nchez} S.~F.,  2011, \mn@doi [\mnras] {10.1111/j.1365-2966.2011.18870.x}, \href {https://ui.adsabs.harvard.edu/abs/2011MNRAS.415.2439R} {415, 2439}

\bibitem[\protect\citeauthoryear{{S{\'a}nchez-Menguiano} et~al.,}{{S{\'a}nchez-Menguiano} et~al.}{2016}]{2016ApJ...830L..40S}
{S{\'a}nchez-Menguiano} L.,  et~al., 2016, \mn@doi [\apjl] {10.3847/2041-8205/830/2/L40}, \href {https://ui.adsabs.harvard.edu/abs/2016ApJ...830L..40S} {830, L40}

\bibitem[\protect\citeauthoryear{{S{\'a}nchez-Menguiano} et~al.,}{{S{\'a}nchez-Menguiano} et~al.}{2018}]{2018A&A...609A.119S}
{S{\'a}nchez-Menguiano} L.,  et~al., 2018, \mn@doi [\aap] {10.1051/0004-6361/201731486}, \href {https://ui.adsabs.harvard.edu/abs/2018A&A...609A.119S} {609, A119}

\bibitem[\protect\citeauthoryear{{S{\'a}nchez} et~al.,}{{S{\'a}nchez} et~al.}{2012}]{2012A&A...538A...8S}
{S{\'a}nchez} S.~F.,  et~al., 2012, \mn@doi [\aap] {10.1051/0004-6361/201117353}, \href {https://ui.adsabs.harvard.edu/abs/2012A&A...538A...8S} {538, A8}

\bibitem[\protect\citeauthoryear{{S{\'a}nchez}, {Walcher}, {Lopez-Cob{\'a}}, {Barrera-Ballesteros}, {Mej{\'\i}a-Narv{\'a}ez}, {Espinosa-Ponce}  \& {Camps-Fari{\~n}a}}{{S{\'a}nchez} et~al.}{2021}]{Sanchez_2021}
{S{\'a}nchez} S.~F.,  {Walcher} C.~J.,  {Lopez-Cob{\'a}} C.,  {Barrera-Ballesteros} J.~K.,  {Mej{\'\i}a-Narv{\'a}ez} A.,  {Espinosa-Ponce} C.,   {Camps-Fari{\~n}a} A.,  2021, \mn@doi [\rmxaa] {10.22201/ia.01851101p.2021.57.01.01}, \href {https://ui.adsabs.harvard.edu/abs/2021RMxAA..57....3S} {57, 3}

\bibitem[\protect\citeauthoryear{{Sharda}, {Krumholz}, {Wisnioski}, {Forbes}, {Federrath}  \& {Acharyya}}{{Sharda} et~al.}{2021}]{2021MNRAS.502.5935S}
{Sharda} P.,  {Krumholz} M.~R.,  {Wisnioski} E.,  {Forbes} J.~C.,  {Federrath} C.,   {Acharyya} A.,  2021, \mn@doi [\mnras] {10.1093/mnras/stab252}, \href {https://ui.adsabs.harvard.edu/abs/2021MNRAS.502.5935S} {502, 5935}

\bibitem[\protect\citeauthoryear{{Smith} et~al.,}{{Smith} et~al.}{2017}]{2017MNRAS.466.2217S}
{Smith} B.~D.,  et~al., 2017, \mn@doi [\mnras] {10.1093/mnras/stw3291}, \href {https://ui.adsabs.harvard.edu/abs/2017MNRAS.466.2217S} {466, 2217}

\bibitem[\protect\citeauthoryear{{Sukhbold}, {Ertl}, {Woosley}, {Brown}  \& {Janka}}{{Sukhbold} et~al.}{2016}]{2016ApJ...821...38S}
{Sukhbold} T.,  {Ertl} T.,  {Woosley} S.~E.,  {Brown} J.~M.,   {Janka} H.~T.,  2016, \mn@doi [\apj] {10.3847/0004-637X/821/1/38}, \href {https://ui.adsabs.harvard.edu/abs/2016ApJ...821...38S} {821, 38}

\bibitem[\protect\citeauthoryear{{Ting} \& {Weinberg}}{{Ting} \& {Weinberg}}{2022}]{Ting22a}
{Ting} Y.-S.,  {Weinberg} D.~H.,  2022, \mn@doi [\apj] {10.3847/1538-4357/ac5023}, \href {https://ui.adsabs.harvard.edu/abs/2022ApJ...927..209T} {927, 209}

\bibitem[\protect\citeauthoryear{{Tinsley}}{{Tinsley}}{1980}]{Tinsley_1980}
{Tinsley} B.~M.,  1980, \mn@doi [\fcp] {10.48550/arXiv.2203.02041}, \href {https://ui.adsabs.harvard.edu/abs/1980FCPh....5..287T} {5, 287}

\bibitem[\protect\citeauthoryear{{Tissera}, {Rosas-Guevara}, {Sillero}, {Pedrosa}, {Theuns}  \& {Bignone}}{{Tissera} et~al.}{2022}]{2022MNRAS.511.1667T}
{Tissera} P.~B.,  {Rosas-Guevara} Y.,  {Sillero} E.,  {Pedrosa} S.~E.,  {Theuns} T.,   {Bignone} L.,  2022, \mn@doi [\mnras] {10.1093/mnras/stab3644}, \href {https://ui.adsabs.harvard.edu/abs/2022MNRAS.511.1667T} {511, 1667}

\bibitem[\protect\citeauthoryear{{Tremonti} et~al.,}{{Tremonti} et~al.}{2004}]{Tremonti04}
{Tremonti} C.~A.,  et~al., 2004, \mn@doi [\apj] {10.1086/423264}, \href {https://ui.adsabs.harvard.edu/abs/2004ApJ...613..898T} {613, 898}

\bibitem[\protect\citeauthoryear{{Turk}, {Smith}, {Oishi}, {Skory}, {Skillman}, {Abel}  \& {Norman}}{{Turk} et~al.}{2011}]{2011ApJS..192....9T}
{Turk} M.~J.,  {Smith} B.~D.,  {Oishi} J.~S.,  {Skory} S.,  {Skillman} S.~W.,  {Abel} T.,   {Norman} M.~L.,  2011, \mn@doi [\apjs] {10.1088/0067-0049/192/1/9}, \href {https://ui.adsabs.harvard.edu/abs/2011ApJS..192....9T} {192, 9}

\bibitem[\protect\citeauthoryear{{Weinberg} et~al.,}{{Weinberg} et~al.}{2019}]{Weinberg19a}
{Weinberg} D.~H.,  et~al., 2019, \mn@doi [\apj] {10.3847/1538-4357/ab07c7}, \href {https://ui.adsabs.harvard.edu/abs/2019ApJ...874..102W} {874, 102}

\bibitem[\protect\citeauthoryear{{Weinberg} et~al.,}{{Weinberg} et~al.}{2022}]{Weinberg22a}
{Weinberg} D.~H.,  et~al., 2022, \mn@doi [\apjs] {10.3847/1538-4365/ac6028}, \href {https://ui.adsabs.harvard.edu/abs/2022ApJS..260...32W} {260, 32}

\bibitem[\protect\citeauthoryear{{Wenger}, {Balser}, {Anderson}  \& {Bania}}{{Wenger} et~al.}{2019}]{2019ApJ...887..114W}
{Wenger} T.~V.,  {Balser} D.~S.,  {Anderson} L.~D.,   {Bania} T.~M.,  2019, \mn@doi [\apj] {10.3847/1538-4357/ab53d3}, \href {https://ui.adsabs.harvard.edu/abs/2019ApJ...887..114W} {887, 114}

\bibitem[\protect\citeauthoryear{{Wibking} \& {Krumholz}}{{Wibking} \& {Krumholz}}{2023}]{WK23}
{Wibking} B.~D.,  {Krumholz} M.~R.,  2023, \mn@doi [\mnras] {10.1093/mnras/stac2648}, \href {https://ui.adsabs.harvard.edu/abs/2023MNRAS.521.5972W} {521, 5972}

\bibitem[\protect\citeauthoryear{{Williams} et~al.,}{{Williams} et~al.}{2022}]{2022MNRAS.509.1303W}
{Williams} T.~G.,  et~al., 2022, \mn@doi [\mnras] {10.1093/mnras/stab3082}, \href {https://ui.adsabs.harvard.edu/abs/2022MNRAS.509.1303W} {509, 1303}

\bibitem[\protect\citeauthoryear{Yang \& Krumholz}{Yang \& Krumholz}{2012}]{Yang_2012}
Yang C.-C.,  Krumholz M.,  2012, \mn@doi [The Astrophysical Journal] {10.1088/0004-637X/758/1/48}, 758, 48

\bibitem[\protect\citeauthoryear{{da Silva}, {Fumagalli}  \& {Krumholz}}{{da Silva} et~al.}{2012}]{2012ApJ...745..145D}
{da Silva} R.~L.,  {Fumagalli} M.,   {Krumholz} M.,  2012, \mn@doi [\apj] {10.1088/0004-637X/745/2/145}, \href {https://ui.adsabs.harvard.edu/abs/2012ApJ...745..145D} {745, 145}

\bibitem[\protect\citeauthoryear{{de Avillez} \& {Mac Low}}{{de Avillez} \& {Mac Low}}{2002}]{2002ApJ...581.1047D}
{de Avillez} M.~A.,  {Mac Low} M.-M.,  2002, \mn@doi [\apj] {10.1086/344256}, \href {https://ui.adsabs.harvard.edu/abs/2002ApJ...581.1047D} {581, 1047}

\makeatother
\end{thebibliography}

%\begin{thebibliography}{99}
%\end{thebibliography}

%%%%%%%%%%%%%%%%%%%%%%%%%%%%%%%%%%%%%%%%%%%%%%%%%%

%%%%%%%%%%%%%%%%% APPENDICES %%%%%%%%%%%%%%%%%%%%%

%\appendix



%%%%%%%%%%%%%%%%%%%%%%%%%%%%%%%%%%%%%%%%%%%%%%%%%%

% Don't change these lines
\bsp	% typesetting comment
\label{lastpage}
\end{document}